\documentclass[12pt]{article}
\usepackage{graphicx}
\usepackage{mathtools}
\usepackage{latexsym}

\newcommand{\be}{\begin{equation}}
\newcommand{\ee}{\end{equation}}
\newcommand{\bea}{\begin{eqnarray}}
\newcommand{\eea}{\end{eqnarray}}

\begin{document}
\begin{titlepage}

\begin{flushright}
{\today}
\end{flushright}
\vspace{1in}

\begin{center}
\Large
{\bf On Production of Excited Kaluza-Klein States in Large Radius Compactification Scenario  }
\end{center}

\vspace{.2in}

\normalsize
\begin{center}
Jnanadeva Maharana \footnote{Indian National Science Academy Honorary Scientist at Institute of Physics, Bhubaneswar. \\ e-mail: maharana$@$iopb.res.in} \\

\end{center}

\normalsize

\begin{center}
 {\em Institute of Physics \\
Bhubaneswar-751005, India \\
NISER, Bhubaneswar, India } \\    

\end{center}

\vspace{.2in}

\baselineskip=24pt

\begin{abstract}
Production of exotic states at LHC is considered in the large
radius compactification scenario. We envisage a five dimensional theory
for a scalar field in five dimensional flat spacetime. It is compactified
on a circle, $S^1$, with  radius, $R$. The radius is assumed to be in TeV
scale appealing to LRC hypothesis. The production of Kaluza-Klein states
whose masses lie in the vicinity of TeV range is considered. Instead
of appealing to any specific model, bounds on inelastic cross sections
and near forward differental cross section are derived from the
Lehmann-Symanzik-Zimmermann (LSZ) formulation. We consider decompactified
theory should compactification radius be large enough to unravel the fifth
spacial dimension in LHC energy scale. Bounds on cross sections are also
derived for this scenario. We present bounds on inclusive cross
sections for reactions like $a+b\rightarrow c+X$, X being unobserved
states. We plot the bounds as a function of energy and propose that these
bounds might be useful for search   of exotic states in 
LHC experiments like ATLAS and CMS.

\end{abstract}

\vspace{.5in}

\end{titlepage}
\section{Introduction  }

The standard model, $SU(3)_C\otimes SU(2)_L\otimes U(1)$, has been tested to 
a great degree of precision. However, there are reasons to believe
that the standard model is not complete and there are several important 
issues which cannot be resolved within the frameworks of the model.
The construction of grand unified theory (GUT) is an attempt to unify
strong, weak and electromagnetic interaction with gauge invariance and 
spontaneously broken symmetry mechanism as  underlying cardinal principles. 
There are hints from experimental observations that
an underlying GUT might be present; however in order to test
some of the predictions of GUT we do not have accelerators of such high
energies. Moreover, other nonaccelerator based tests like instability of
proton have not yielded any definite results so far.
Therefore, there are no clinching evidences in favor of 
GUT paradigm.  Furthermore, gravitational interaction is not incorporated in
GUT. Therefore, a unified theory of four fundamental interaction must 
incorporate gravitational interaction. Superstring theory
offers the prospect of unifying fundamental forces of Nature. 
Consequences of quantum theory of
gravity are expected to be addressed from string theory perspective. 
 A consistent 
quantum description of superstring string  requires that it must live
in ten spacetime dimensions. We have no experimental evidence of 
ten dimensional spacetime manifold so far. In this context, the ideas of Kaluza \cite{kaluza} and
Klein  \cite{klein} (KK) have been revived. A comprehensive review and 
historical developments of the idea of theories in higher spacetime
dimensions and their influence on developments of diverse proposals 
in supersymmetry, supergravity theories as well as in string theories
are presented in the book  edited by Appelquist, Chodos and Freund 
\cite{appelquist}. 
This book contains  collection of large number of important articles  
on the subject of 
higher dimensional theories and various schemes of compactifications. 
 Kaluza and Klein had set out to  provide a unified description of 
electromagnetism and gravitation following  Einstein's theory of gravitation 
whose foundation lies in general theory of relativity. It is worthwhile to note that the idea of describing physics in higher
dimensions predates  works  of Kaluza and Klein \cite{predate1,predate2} 
which was envisaged before Einstein expounded the
 general theory of relativity. 
  Their  endeavors were to unify Maxwell's theory with four dimensional 
continuum. Kaluza and Klein considered a theory of gravitation 
in $D=5$ described
by Einstein-Hilbert action. They compactified one spatial dimension on a circle, $S^1$, of radius $R$ and argued that $R<<1$ so that the
experiments of that era could not probe that length scale. The resulting 
four dimensional action described Einstein's theory of relativity, 
with a massless
spin two graviton,  an Abelian
gauge field  and a massless scalar field  appearing as zero modes.
These fields interacted through exchange of massless particles 
(so called zero modes. important for long range forces ) which arise
as a consequence of KK compactification.  Moreover, the resulting 4-D action 
also contained tower of massive spin two,
spin one and
spin zero field.  
Besides these long range forces,
there are short range forces due to exchange of massive KK fields. 
 Moreover, the resulting 4-D action also contained tower of massive spin two, 
spin one and
spin zero field. However, the proposal of KK was not successful to describe 
physical phenomena of that era.  The Kaluza-Klein paradigm was 
examined by several authors; notably among them are Schr\"odinger \cite{erwin},  Jordan \cite{jordan},  Bergmann and Einstein \cite{be},
 Pauli \cite{pauli} and  Bergmann \cite{bergmann}. 
 The idea was revived when consistent
supersymmetric and supergravity theories were constructed in 
higher dimensions, $D>4$. In an important paper, Scherk and Schwarz \cite{ssch}
presented the   procedure for compactification of higher dimensional  
theories through  a systematic  prescription of dimensional reduction.
This technique has been widely used in present times. Furthermore, string 
theories offered the prospect of unification
of fundamental forces since it was possible to construct theories which had 
nonabelian massless gauge bosons, chiral fermions and admitted those
gauge groups which render the theory to be free from anomalies. 
The construction of heterotic string in $D=10$ had several desirable
attributes and led to vigorous activities to obtain four dimensional 
string effective action though compactification schemes. Thus
there were attempt to derive realistic string effective actions in
order to investigate phenomenological implications of string theories.
 It was generally accepted that the compactification
scale will be in the vicinity of Planck scale or scale of grand unified 
theories (GUT). We recall that the string tension is of the order of Planck scale and
therefore, massive excited stringy states will have mass spacings 
 of the order of that scale. If the compactification radius corresponds to 
 the scale mentioned above then
the KK zero modes are to correspond with the known particle spectrum and 
KK excitations will have masses proportional to inverse
of compactification radius as will be discussed in the sequel. In this optics,
 there seems to have a slim chance of experimentally observing  the
excited states of strings in foreseeable future.  Thus if the 
compactification scale is at Planck regime or in the GUT regime, the chances of
experimentally observing stringy states, in $D=4$,  
even in distant  future is  
almost hopeless. We do not have accelerator technology to create Planck
scale collision energies. 

An alternative scenario, for low scale compactification, envisages a low scale 
(i.e. large radius) compactification of higher dimensional ($D>4$) 
theories \cite{antoniadis, others}. Subsequently,
an attractive proposal for large radius compactification was introduced which 
incorporated several interesting novel features \cite{add1,add2,aadd}.
 There is strong basis to pursue
phenomenological analysis of this proposal. We mention in passing that
the large radius compactification proposal is not  specifically
intended for string theory effective actions. A theory which is
proposed in higher dimensions, $D>4$, when compactified to 'physical'
lower dimension, $D=4$, is amenable to this paradigm. Moreover, there are
 hopes 
that the predictions could be subjected to experimental tests. It is proposed,
in this scenario, that scale of compactification could be in the
vicinity of a few TeV and there are hopes  that predictions of this 
proposal could be tested at LHC energies. Indeed, 
ATLAS and CMS are searching for such particles
which might be produced at LHC. These two experiments have obtained lower limits on mass of such 'exotic' particles which range from $2$ TeV
to $6$ TeV \cite{Tev1,Tev2}.  Of course,  the experimental searches to obtain limits on masses of excited states
are  guided by inputs from  various phenomenological models.

Instead of proposing another  model for production of spectrum of particles 
arising from large-radius-compactification scheme, we derive bounds on
production cross section of the so called exotic KK states. We work in the 
framework of general field theories with minimum inputs from models.
These bounds are derived in a nonperturbative framework; just as the
Froissart bound on hadronic total cross section, $\sigma_t$, does 
not appeal to perturbation theory. We shall dwell on this aspect, in detail,
in subsequent sections.   We mention,
in passing, that we are handicapped by the fact that there is no 
experimental data to be used as inputs.  
Therefore,
these bounds are derived from first principles of a theory. Furthermore, 
there are a host of models of large-extra-dimensions which 
are built with the original paradigm of \cite{add1,add2,aadd}. The topic has 
been extensively reviewed in the forms of lectures
at various schools and workshops 
\cite{antrev,luest,acc,gs,r1,r2,r3,r4,r5,r6,r7,r8,r9}.  

We envisage a 
five dimensional theory for the sake of simplicity. The
arguments presented here can be extended to a theory in $D>4$ dimensions 
since the analyticity properties have been investigated in detail by us
\cite{jmjmp}. We are also aware that higher dimensional theories, beyond
five dimensions might face difficulties while compactified in large
radius compactification scenario from phenomenological considerations. However,
we present our investigation from a rigorous and new perspective so
that insights gained from this work might be useful.
We add the following remarks to clarify some statement made in this
article from time to time. We shall use terms such as strong bound (result) 
and 
weak  bound (result) on and off. These two terms are to be understood in the
following sense. A strong bound (result) means that given a set of basic
assumptions or axioms if the energy dependence of a bound (generally power
of $log s$ is less compared to another bound (which is derived under
same set of assumptions (axioms) has a higher power of $log s$ then 
we designate former to be a stronger bound. In our case, we work
in the frameworks of LSZ axioms; therefore, with above definition of stronger
and weaker bounds we shall be able to distinguish quality of bounds
(i.e. strong or weak).

We consider production of Kaluza-Klein states in high energy collision of
protons at LHC. The process under considerations is:
$p+p\rightarrow K_1+K_2$ where $K_1$ and $K_2$ are excited KK states.
They carry baryon number and also KK charges. The global KK charges
are conserved as will discussed later. Therefore, if $K_1$ is assigned
baryon number $B=1$ and KK charge $q_n$ then $K_2$ carries baryon number
$B=1$ and KK charge $-q_n$. Moreover, KK charges are quantized and are
proportional to an integer. If $R$ is the radius of compactification of the
circle then the mass formula a KK state is 
$m_n^2=m_0^2+ ({{n}\over R})^2, n\in{\bf Z}$; here $m_0$ is mass of the 
zero mode (see more discussions later). 
All these aspects will be dealt with in detail in the next section. Obviously,
lowest mass KK 'baryon' is heavier than proton. Correspondingly,
the s-channel threshold is $s=(m_p+m_1)^2$ where $m_1$ is the mass of first
excited KK state (carries $B=1$ and $n=1$). 
 Note that $m_p$ is the proton mass which corresponds to
mass of zero mode in KK compactification. 
The t-channel threshold is also important in that
its location plays an important role to write down fixed-$t$ dispersion 
relations. The threshold is $(m_p+m_{K_1})^2$. We note that the
production process, $p+p\rightarrow K_1+K_2$ is an inelastic reaction
where masses of incoming  particles are not same as outgoing particles.
Our attentions will be primarily focused on inelastic reactions.
Let us note a few simple points. 
Notice that the lowest mass KK excitation corresponds to $n=\pm 1$ i.e. 
one of the states, say $K_1$ carries quantum numbers $B=1, q_n=1$ whereas
$K_2$ has quantum numbers $B=1,q_n=-1$. This reaction is consistent with
all conservation laws (all the particles are strongly interacting).
We shall discuss
the analyticity properties of inelastic reactions, generically denoted
as $a+b\rightarrow c+d$ in subsequent sections. The purpose of this articles is twofold. (i) We envisage a situation where 
one spatial dimension is compactified on a circle; it is the so called $S^1$ 
compactification.
As we shall describe in detail, in order to study analyticity properties of 
such a theory it is necessary to construct the Hilbert space where not only the zero
modes define a state vector but state vectors associate with KK excitations 
are to be included. It is essential to derive spectral representation 
of the amplitude where 
the complete set of states consists of the states associated with the zero mode sector and the  set of states associated with the KK towers.
We shall present arguments on the spectral representation in some details. 
Moreover, the case of inelastic scattering requires some extra 
care due to additional
technical problem. (ii) We also consider the situation when the 
theory has no compact dimension i.e. noncompact spatial coordinates. It is essentially
the problem of investigating a $D=5$ field theory and deriving  analyticity properties of scattering amplitude. This aspect has been comprehensively
investigated by us \cite{jmjmp}. For the case at hand, we have to deal with four point amplitude. Here basis functions for partial wave expansion are
the Geggenbauer polynomials.

We consider a simple  model where it has only one type of field in $D=5$ 
to start with. When we compactify this theory on $S^1$ and look
at the spectrum in manifold  
$R^{3,1}\otimes S^1$, it  consists of a field which has the same mass as 
the $D=5$ theory. This corresponds to the zero mode. 
Moreover, there are  KK towers
of  massive fields.  This argument can be extended to the case where $D=5$ 
theory has several species of fields.  In a realistic scenario,
indeed, all the fields of the standard model  (of $ D=4 )$ have their 
counterparts in the five dimensional world. Therefore, from the perspective of
$D=5$ theory, the standard model,  $D=4$ theory, fields are 
zero modes of compactified $D=5$ theory. Naturally, there will be KK modes associated
with species present in $D=5$ theory. Let us consider a single field in the 
$D=5$ theory.  While deriving spectral representation for four point
amplitude, in $D=5$ theory, 
we have to sum over complete set of physical states. 
Moreover, since there is only one field, the four point amplitude 
describes elastic process.
If we have several species of particles in $D=5$ then there will be elastic 
amplitude as well as inelastic amplitude. Therefore, it is important
to study analyticity properties of inelastic reactions as well. 
We shall return to discussions of this subject in later sections.

 We assume that the momenta of the produced exotic particles lie in three 
spatial, physical,  dimensions i.e. $k_i= (k^0_i, {\bf k}_i),~i=1,2$,
where ${\bf k_1}$ and $ \bf{k_2}$ are three 
momenta of two outgoing particles. The energies and momenta of the exotic particle can be measured in LHC experiments.
This is our choice of kinematical configurations.
 We study production of states  in LHC. There are two possibilities which we have alluded to above. The excited particles
 produced in case of $S^1$ compactification are henceforth referred 
to as Kaluza-Klein (KK) states, which appear 
in large-radius-compactification models,
in collisions at LHC. We derive bounds on total production cross section,
 bounds on differential 
cross sections and bounds on
inclusive cross sections based on principles of general local quantum field 
theories (QFT). Therefore, our efforts are to minimize the inputs from
specific models as much as possible. There are
 certain differences between existing rigorous bounds derived 
from axiomatic field theories, in $D=4$,
and the problem under consideration. There are various scenarios, 
adopted from different phenomenological perspectives, within LRC paradigm. 
The references cited here, especially several of the TASI lectures, 
deal in detail, attributes of LRC models  and present  
predictions from diverse  perspectives.
We have made the following choice and it is a simple extension 
of the standard model as far as the underlying assumptions are concerned. 
 It is 
called universal extra dimension (UED) paradigm (see \cite{r7} 
for a review of this idea).
  It is proposed that the four dimensional standard model (SM) is promoted to a 
  $\hat D$ dimensional
theory where ${\hat D}=4+n$,  $n$ is the number of extra spatial dimensions. 
Therefore, the fundamental constituents of
the standard model such as  gauge bosons, fermions and  scalars are promoted to fields in $4+n$ dimensions. The basic building blocks of the
4-dimensional world appear as zero modes of the compactified theory. 
Therefore, additional KK states appear  after the $4+n$ dimensional 
theory is
compactified.   It is assumed that extra dimensions are flat and so is 
our four dimensional world where SM is experimentally tested. 
This approach has several advantages from phenomenological point of view 
and it has also shortcoming as is the case with any model in the LCR scenario. 
We have chosen this from 
certain advantages it offers from our perspective as will be 
elaborated soon. 
In the context of model buildings, the practitioners 
incorporate extra assumptions
so that perturbative corrections coming from states arising from 
compactifications do not upset results of $D=4$ theories at tree level. 
These aspects
(for that matter those assumptions) are not needed here.  
In this scheme some of their inbuilt  results are of interests to us. 
Parity is an exact symmetry.
A single, first excited KK state $(n=1)$ cannot be produced in isolation.  
The lightest state, $n=1$, is stable. However, it is not very obvious that 
 $n=1$ KK quantum  number be an absolutely conserved  
charge since it is not result  
of an exact local symmetry (see discussions in \cite{jm1,jm2}).
We remark that the toroidal compactification of a higher dimensional theory 
(where ${ D}$ dimensional theory is flat) satisfies LSZ axioms
\cite{jmjmp}. Moreover, the Hilbert space structure of flat space $D=4$ theory 
derived from a $D=5$ flat space theory has been discussed in \cite{jm1,jm2} in 
detail. Furthermore, toroidal compactification of a higher dimensional 
theory to $D=4$   theory poses no additional difficulties. In view of preceding
remarks we consider   the essential aspects of UED. 
We  focus on the scalar sector. 
We intend to explore another scenario where the extra dimension is 
decompactified, already, at LHC energy. Thus states of the five dimensional (decompactified)
model will be produced. A systematic analysis is quite desirable for following reasons. First the bound on total cross sections, the generalized Froissart bound,
is different for $D>4$. This case has been rigorously studied by 
us \cite{jmjmp}. Furthermore, when we derive spectral representations
for absorptive part of four point amplitude in the compactfied, $D=4$, 
theory, the KK states are introduced as physical intermediate states. 
This aspect
is discussed in the next section. On the other hand, in a decompactified theory (here $D=5$ theory), the intermediate states appearing in the
spectral representation are physical states. We work in the LSZ formulation. Therefore, all the parameters such as mass, charge etc are the physical
observables. In this framework, all quantum effects are accounted for. For example let us consider $\pi N$ scattering. The proof of Froissart bound
does not rely on any perturbation theoretic frame work. Similarly, derivation on properties of diffraction peak such as shrinkage of width of diffraction peak
and location of minimum of differential cross section as a function of $t$ are derived in this framework. Therefore, LSZ formulation, when applied to
decompactified theory will capture effectively the quantum effects without 
resorting to perturbation theoretic arguments\footnote{I thank 
A. P. Balachandran for discussions on this point. He emphasised that
Froissart bound and its subsequent refinement by Martin were accomplished
long before QCD was proposed (and established) as the theory of
strong interactions. Indeed, models to study strong interactions in
the field theoretic framework were not proved to be renormalizable, like
QED, on various grounds.
 Therefore, establishing bounds for theories
in higher dimensions ( with compact or noncompact coordinates) might play
an useful role in future. }. 
The presence of spinning particles do not pose any serious problem
\cite{book1,book2}\\.

 The paper is organized as follows. We shall present the general structure of the compactified theory in the next section, Section 2.
  We briefly summarize the contents of \cite{jm1,jm2}
which will be utilized in this work. We refer the interested reader to go through the details in the above mentioned papers.  The starting point
is to consider a massive scalar field theory in ${\hat D}=5$ dimensional flat space with Lorentzian signature.  It was shown in \cite{jmjmp} that LSZ procedure
is applicable to study analyticity properties for higher dimensional theories, 
$D>4$. Notice that the asymptotic states i.e. 'in' and 'out' states
are defined in LSZ formulation and this concept is not special to 
four dimensions. Another important point is to be kept in mind is that
in LSZ approach the only the physical parameters such as mass of particles 
and physically measured coupling constants are used. In essence, as is
evident from LSZ axioms, all results are derived without addressing 
to perturbation theory. 

  We derive results
on analyticity properties of scattering amplitude for inelastic
two body reactions in Section 3. There have
been previous attempts to study analyticity of inelastic reactions in
the past \cite{russian1,russian2}. However, these authors were (probably)
unaware of the
intricacies associated with unequal mass scattering in deriving analyticity
domains in the $t$-plane.  As we shall present, in the sequel,
identification
of the holomorphic domain in $t$-plane is to be treated with some care
since the physical energy threshold is not identical to elastic threshold.
We have invoked
rigorous theorems from axiomatic field theory to argue that the amplitude
of $s$-channel reaction can be analytically continued to the $u$-channel
reaction.
Some of these delicate points were not addressed in the past
\cite{russian1,russian2}. Next we discuss the existences of Lehmann ellipses
since
these results were derived earlier and what is the analog of Martin'e ellipse.
Subsequently, a fixed-$t$ dispersion relation is written down. 
We devote Section 4
 to deriving bounds on various  cross sections alluded to above.
We have employed refined technique of Singh and Roy \cite{sr70} as
well as Einhorn and Blankenbecler \cite{eb70}  to
improve results of \cite{russian1,russian2} where the prefactors in
their bounds appear as arbitrary constants. This arbitrariness could be
removed by
invoking more refined procedures which are outlined in this section.
 Section 5 deals with
case when the $5^{th}$ dimension is decompactified. There are simplifications
in kinematical configurations for four point function. It is relatively
simple to derive known results of $D=4$ theory for the $D=5$
theory while considering
the four point functions. The proof of edge-of-the wedge theorem of $D=4$
theory is practically carried over for 4-point function of $D=5$ theory. 
This leads,
{\it de facto},
to proof of crossing which is crucial to write fixed $t$ dispersion relation.
 We do not repeat the proof of our derivation \cite{jmjmp} here. However,
we present
the essential arguments. Once we are able to write fixed $t$ dispersion
relation for inelastic amplitude we have proved the analyticity properties
in this case.
 We discuss the bounds on inclusive reaction which are obtained by adopting
techniques
used to derive bounds on inelastic processes.
 
 The last section, Section 6, contains summary of our
results and discussions. It is important to note that if the exotic particles
are produced  in LHC energy regime then new thresholds will open up. 
If such is the case then the effects of the presence of new threshold
will affect the analyticity properties of scattering amplitude.  Indeed,
a precise measurement of real part of the scattering amplitude ( in other words
measurement ratio of real to imaginary part) 
at energies below the threshold will
signal the possibility of production of exotic particles. We shall
discuss this aspect in discussion section.  

\bigskip

\section {Properties of inelastic scattering amplitude for $S^1$ compactified theory}

We consider a scalar field, 
${\hat\phi}({\hat x}), {\hat x}^{\hat \mu}=0,1,2,3,4$ in  a five dimensional
flat space.  
It is convenient to  decompose the
five dimensional spacetime coordinates, 
${\hat x}^{\hat\mu}$  as follows for later conveniences:
\be
 {\hat x}^{\hat\mu}=(x^{\mu}, y)
  \ee
  where $x^{\mu}$ are flat  space coordinates, $\mu=0,1,2,3; y$ is the 
  compact coordinate on $S^1$ with periodicity
   $y+2\pi R = y$, $R$ being the radius of $S^1$.  
   As a consequence of $S^1$ compactification the resulting spacetime 
manifold is  
$R^{3,1}\otimes S^1$. 
  We  remind that asymptotic 'in' and 'out' fields satisfy free field equation 
in 5-dimensions before we discuss consequences of compactification. The
  equation of motion, for $D=5$ theory, is 
  $[{\Box}_5+m_0^2]{\hat\phi}^{in,out}({\hat x})=0$, $m_0$ being the 
mass of the scalar field. 
  We expand the field, 
${\hat \phi}(x,y)$, in a Fourier series in $y$-variable after implementing
$S^1$ compactification
  \bea
  \label{kk1}
  {\hat\phi}^{in,out}({\hat x})={\hat\phi}^{in,out}(x,y)=\phi^{in,out}_0(x)+
\sum_{n=-\infty, n\ne 0}^{+n=\infty}\phi^{in,out}_n(x)e^{{{in y}\over{ R}}}
  \eea
  where $\phi^{in,out}_0(x)$, the  zero mode, has no $y$-dependence. 
The terms in rest of the series  (\ref{kk1})
  are periodic in  $y$. The  five dimensional Laplacian, ${\Box}_5$, 
is decomposed  as sum two operators: 
    $\Box_4$ and  ${{\partial}\over{\partial y^2}}$ . The equation 
of motion is
  \be
  \label{kk2}
  [\Box_4 - {{\partial}\over{\partial y^2}}+m_n^2]\phi^{in,out}_n(x,y)=0
  \ee
  where $\phi^{in,out}_n(x,y)=\phi_n^{in,out}e^{{{in y}\over{ R}}}$ and $n=0$ 
term has no $y$-dependence being denoted as $\phi_0(x)$; from now on
  $\Box_4=\Box$.
  It follows from equations of motion that $m_n^2=m_o^2+{{n^2}\over{R^2}}$. 
Therefore, there is a tower of massive  states. The momentum along
  $5^{th}$ direction is quantized, $p_4=q_n={{n}\over R}$.
  $q_n$  is an additive conserved quantum number. We continue to call it  
Kaluza-Klein (KK) charge and note that we have considered a flat space theory; 
   there is no gravitational interaction in the five dimensional theory
   we have adopted.  
For the interacting field ${\hat\phi}({\hat x})$
\be
   \label{kk2x}
   {\hat\phi}({\hat x})={\hat\phi}(x,y)=\phi_0(x)+\sum_{n=-\infty, n\ne 0}^{n=+\infty}\phi_n(x)e^{{{iny}\over R}}
   \ee
   The equation of motion for the interacting fields has a source terms on 
the $r.h.s$. Naturally, the source current is  expanded in Fourier series 
   as is the expansion (\ref{kk2x}). Each field $\phi_n(x)$ will have a 
current, $J_n(x)$ associated with it as source
   current and the expansion is
   \be
   \label{kk2a}
   {\hat j}(x,y) =j_0(x)+\sum_{n=-\infty, n\ne 0}^{n=+\infty}J_n(x)e^{i{ny/R}}
   \ee
   The set of currents,  $\{J_n(x)  \}$, are the source currents associated 
with the tower of interacting fields
   $\{ \phi_n(x) \}$. These fields carry the discrete KK charge, $q_n$. 
Therefore, $J_n(x)$ also carries the
   same KK charge. This is to be kept in mind while going through LSZ 
reductions to write down the amplitudes and Greens functions. 
  The zero mode, $\phi^{in,out}_0$, creates its own    
Fock space. Similarly, each of the KK fields, $\phi^{in,out}_n(x)$, 
creates  its own  Fock space. 
  We present some illustrative examples below. A state with spatial momentum, 
${\bf p}$, energy, $p_0$ and discrete momentum $q_n$  ( KK state) is created
  by
  \be
   \label{kk3}
   A^{\dagger}({\bf p},q_n)|0>=|p,q_n>,~~p_0>0 ~~i.e. ~p_{\mu}\in V^+
   \ee
  Our starting point was a 5-dimensional theory of a scalar field.   
  After  $S^1$ compactification, the geometry is  $R^{3,1}\otimes S^1$ and 
we have discussed the spectrum of the compactied theory. 
  A massive field of mass $m_0$, the mass of zero mode, and a the KK tower 
characterized by masses and 'charges' $\{(m_n, q_n)\}$ are in the spectrum.
  \\
  The Hilbert space associated with the five dimensional theory 
is ${\hat {\cal H}}$. On compactification, the resulting theory, 
$R^{3,1}\otimes S^1$, 
   leads to decomposition of the original
  Hilbert space into direct sum of Hilbert spaces each of which is
 characterized by its KK charge $q_n$.
  \be
  \label{kk4}
  {\hat{\cal H}}=\sum \oplus {\cal H}_n
  \ee
  Thus ${\cal H}_0$ is the Hilbert space constructed from $\phi_0^{in,out}$ 
with charge $q_{n}=0$ and it is   built by the actions of the creation 
operators 
  $\{ a^{\dagger}({\bf k}) \}$  acting on the vacuum. The resulting states 
span ${\cal  H}_0$. 
   A single particle state is  $a^{\dagger}({\bf k})|0>=|{\bf k}>$ and 
multiparticle states
  are created using the procedure outlined above. 
   Note the orthogonality relation between two states of different values of 
$q_n$ belonging to two different vector bases, 
  \be
  \label{kk5}
  <{\bf p}, q_{n}|{\bf p}',q_{n'}>=\delta^3({\bf p}-{\bf p'})\delta_{n,n'}
   \ee
  It is assumed that  there are no bound states in the theory.
   \\
  The LSZ formalism can be adopted for the compactified theory. If we keep in 
mind the steps
  introduced above, it is possible to envisage field operators $\phi^{in}_n(x)$
 and $\phi^{out}_n(x)$ for each of the fields for a given $n$.
  Therefore, each Hilbert space, ${\cal H}_n$ will be spanned by the state 
vectors created by operators $a^{\dagger}({\bf k})$, for $n=0$ and
  $A^{\dagger}({\bf p}, q_n)$, for $n\ne 0$.  \\
  {\it Remark}: Note that in (\ref{kk1}) sum over $\{n\}$ runs over positive 
and negative integers. If there is a parity symmetry $y\rightarrow -y$
  under which the field is invariant we can reduce the sum to positive 
$n$ only leading to $cos{{n\theta}\over R}$. On the other hand if  the field 
is odd
  under $y\rightarrow -y$ we have $sine$-function. We make no such assumptions 
here. 

 We considered three types of scattering process: (a) scattering of zero modes,
 $\phi_0+\phi_0\rightarrow \phi_0+\phi_0$. It is the conventional scattering
 of two massive scalars. (b) A zero mode, $\phi_0 $ scattering with a 
KK excitation with charge $q_n$. This is an elastic scattering of zero 
KK charge particle
 with nonzero KK charge particle. The processes (a) and (b) are treated 
without complications
 adopting the well known precedures. 
(c) The reaction $\phi_m+\phi_n\rightarrow \phi_{m'}+\phi_{n'}$
 deserves attention as KK charge conservation is required i.e. 
$q_m+q_n=q_{m'}+q_{n'}$. We had considered elastic reaction 
since our goal was to prove
 dispersion relation in order to resolve problems posed by Khuri's  
\cite{khuri}analysis in potential scattering. The study of analyticity 
of elastic process in  (c) i.e $\phi_+\phi_n\rightarrow \phi_m+\phi_n$
 required careful investigations  since the intermediates appear from the 
entire Hilbert space, $\oplus {\cal H}_n$,  as long as all conservation 
laws are 
 satisfied. These issues have been discussed in detail by in references 
\cite{jm1,jm2}. 
 We are not going to repeat those computations in this article. We focus 
on inelastic reactions and  analyze the analyticity properties of 
 corresponding scattering amplitude.  It will be evident that additional 
care should be exercised when we identify domains of holomorphy in $s$ and 
in $t$
 planes.   Moreover, the results presented here on bounds are improvements 
over previous works where due attention was not carefully paid
 in study of certain properties of the domains of holomorphy.

 We investigate,  in this section, two body inelastic scattering amplitude 
for $S^1$ compactified theory. We 
 consider inelastic four point amplitude, $F(s,t)^{ab\rightarrow cd}$,  
where $c$ and $d$ are 'KK' states of nonzero KK charge.
We shall elaborate on this aspect in the sequel. Consequently, we obtain 
bounds on 
$\sigma_t^{ab\rightarrow cd}$, on 
${{d\sigma^{ab\rightarrow cd}}\over{d\Omega}}$
in the forward direction as well as in nonforward direction: 
$\theta_{c.m}\ne 0,\pi$.  The bound on $\sigma_t^{ab\rightarrow cd}(s)$ 
is not as stringent as the
Froissart bound \cite{froissart}, 
$\sigma_t(s)\le {{4\pi}\over{t_0}}[log{{s}\over{s_o}}]^2$ , and we 
shall dwell upon this aspect when we arrive at the bound. 
We also consider
inclusive reaction, $a+b\rightarrow c+X$ where $c$ is the observed KK state 
and X could be a single particle or collection of multiple particle. Feynman
had introduced the notion of inclusive reactions \cite{feynman} in the 
context of hadron collisions and had proposed scaling phenomena based on 
his parton model.
We present bounds on differential cross sections of inclusive reaction in 
the context of production of KK states.

It is important to mention the difficulties, at this juncture,  what one has to 
surmount is deriving rigorous bounds on inelastic processes in general. 
We recall
that Froissart bound is derived from following ingredients, which are 
proved from axiomatic field theory. (i) Analyticity of scattering amplitude 
in the complex
$s$-plane which is also analytic in complex $t$-plane within the Lehmann 
ellipses. We need to  prove existence of Martin ellipse. 
(ii) Crossing symmetry. (iii) Convergence of partial wave expension
inside Lehmann-Martin ellipse. Whereas the results (i) and (ii) 
can be proved for inelastic two body reaction, $a+b\rightarrow c+d$; the 
 existence of
Martin's ellipse is proved from positivity properties of absorptive elastic 
scattering amplitude and positivity properties of elastic
partial wave amplitudes 
$0 \le | f_l(s)^{el}|^2\le {\rm Im}~f_l(s)^{el}\le 1$ (follows from unitarity).
We refer to books and review articles on the study of analyticity properties 
of scattering amplitude from the perspectives of axiomatic field theory 
\cite{book1,book2,book3, fr1,lehm1,sommer,eden,jost,streat,kl,ss,bogo}. 

 In fact the partial wave amplitude of inelastic process do not enjoy 
positivity properties as those of the elastic case.
 Therefore, proof of existence of Martin's ellipse and proof 
holomorphy of scattering amplitude in a complex $t$ domain  requires
some careful analysis. We analyze these issues to identify the domains of 
analyticity of scattering amplitude. Therefore, a fixed-$t$ dispersion relation
is written down only after these issues are addressed and accounted for. 
Recently, we addressed the case of analyticity  of scattering
 amplitude in field theory where one spatial dimension is compactified on 
a circle
of radius $R$, $S^1$ compactification. In fact Khuri  \cite{khuri} had shown, 
in a nonrelativistic potential model with $S^1$ compactification, 
that the forward scattering
amplitude does not satisfy dispersion relation. If so were the case in 
relativistic QFT it would have disastrous consequences. We proved,
starting from LSZ axioms, that the scattering amplitude not only satisfies 
forward dispersion relation but also nonforward dispersion relations are 
satisfied
\cite{jm1,jm2}. The proof was presented for elastic scattering. 
The proof of dispersion relation for inelastic scattering, for a theory 
with $S^1$ compactification,
is not so straightforward as we shall discover.

\bigskip

\section{Production of Particles in Large-Radius-Compactification Proposal  }
 
 We discuss production of particles which belong to   spectrum of states 
 of large-radius-compactifaction (LRC) scenario in the UED paradigm. Our purpose is to derive
 bounds on cross sections without appealing to a specific model. The starting point is to consider an inelastic process $a+b\rightarrow c+d$ where
 $a$ and $b$ are particles accelerated in an accelerator  (they are two colliding protons in LHC experiments) and $c$ and $d$ are two particles
 which belong to the spectrum of particles (charged KK states) in the 
 LRC models with the hypothesis of UED.  
 Existence of such exotic particles is  yet to be confirmed experimentally.
  We refer to them, generically, as Kaluza-Klein (KK)
 states since some of the particle of UED  scenario  belong to 
 KK excitations.  We noted that  KK, exotic states carry additive 
 quantum numbers (charges),
 denoted as $\{ q_n \}$ which are conserved. The KK charges are quantized since they are associated with momenta in the compact directions.
 These additive charges may be viewed to be analogous to baryon number or lepton number. In view of preceding remarks, these particles will be
 produced in pairs and carry equal and opposite KK charges besides other quantum number. For example, if at LHC, in $pp$ collision
 a pair of these exotic particles carrying two units of KK charge each (equal and opposite charges) one will have quantum number of baryon number  and
 two units of KK charge ({ \it this is like strangeness quantum number which is additive}).  The other one will carry baryon number but opposite KK charge
 for total KK charge conservation. Thus one of them might decay to a particle carrying baryon
 number and one unit of KK charge and to a boson with one unit of 
opposite KK charge (if kinetics permit such decays). 
We are not proposing a phenomenological analysis of such a production
 mechanism. We present it,  rather as an illustrative reaction. 
  We have noted earlier that there are lower bounds on masses of particle of LRC theories set by ATLAS and CMS collaborations
 of LHC. 
 
 Our  goal is to examine analyticity properties of the aforementioned scattering amplitude from principles of general field theory which will be presented
 shortly. Notice that we envisage inelastic scattering amplitude which does not satisfy all the properties of elastic scattering amplitude. In particular,
 the elastic partial wave amplitudes, $f_l^{ab\rightarrow ab}(s)$,  satisfy positivity constraints:
  $0\le | f_l^{ab\rightarrow ab}(s)|^2\le Im~f_l^{ab\rightarrow ab}(s)\le 1$,  
whereas inelastic amplitudes do not. Moreover, the domain of analyticity of inelastic amplitudes, in the complex t-plane, especially the proof 
from the existence
of Martin's ellipse depends on ingredients which are proved from the properties enjoyed by  elastic amplitude. Therefore, our first step is to
recapitulate some of the important results on analyticity properties of elastic amplitudes which have been proved in the axiomatic field theoretic framework.
These will be utilized to derive analyticity of inelastic scattering amplitudes.
We shall adopt the convention that a scattering amplitude will have a superscript denoting the reaction and it will be evident from this notational
convention whether we are considering an elastic process or an inelastic 
process.

\bigskip

\noindent{\bf Notations and Kinematical variables}

\bigskip

We consider inelastic reaction
\bea
\label{sec2.1}
a+b\rightarrow c+d
\eea
For the case at hand masses of particles $a$ and $b$ are equal, $m_a=m_b$ and the masses of two outgoing particles, $c$ and $d$, are
also equal, $m_c=m_d$. The initial momenta of a and b are denoted as $p_1$ and $p_2$ whereas final state particles, c and d carry momenta $k_1$ and $k_2$.
The Mandelstam variables are
\bea
\label{sec2.2}
s=(p_1+p_2)^2=(k_1+k_2)^2,~~t=(p_1-k_2)^2,~~u=(p_1-k_1)^2,~~s+t+u=2(m_a^2+m_c^2)
\eea 
 We define ${\bf P}_1(s)$ and ${\bf K}_1(s)$ to be c.m. momenta of particles $(a,b)$ and $(c,d)$ respectively and 
 \bea
 \label{sec2.3}
 P_1^2(s)={1\over 4}(s-4m_a^2)~~~{\rm and}~~~K_1(s)^2={1\over 4}(s-4m_c^2)
 \eea
Let $\theta$ be the scattering angle in the c.m. frame then
\bea
\label{sec2.3}
cos\theta={{{{\bf P}_1(s)^2+{\bf{K}_1(s)^2}}+t}\over{2|{\bf P}_1(s)||{\bf K}_1(s)|}}
\eea  
We also define following kinematical variables of interest
\bea
\label{sec2.4}
({\bf M}_a^2, ~~{\bf M}_b^2, ~~{\bf M}_c^2,~~{\bf M}_d^2)~~{\rm and} ~~({\bf M}_{ab}^2,~~{\bf M}_{ac}^2,~~{\bf M}_{ad}^2,~~{\bf M}_{bc}^2,~~{\bf M}_{bd}^2,~~
{\bf M}_{cd}^2)
\eea
Where ${\bf M}_i^2, i=a,b,c,d$ corresponds to $(mass)^2$ of two or more particle states which carry quantum numbers of particle  $'i'$. Similarly,
${\bf M}_{ij}^2, i,j=a,b,c,d, i\ne j$ corresponds to two or more particle states which carry quantum numbers of both the particles, $'i'$ and $'j'$. This
this the case when we consider various scattering channels. We shall use these definitions in the sequel.

Next, we consider LSZ reduction procedure for four point function: $a+b\rightarrow c+d$. 
We denote the fields associated with outgoing particles as $\phi_n$ and $\phi_{-n}$  so that they carry opposite KK charges; $n$ takes positive or
negative integer values. Note that the fields might carry other quantum numbers. 
 We adopt LSZ \cite{lsz} procedure to derive expression for the 4-point inelastic amplitude. The axioms of LSZ are\\
{\bf A1.} The states of the system are represented in  a
Hilbert space, $ {\cal H}$. All the physical observables are self-adjoint
operators in the Hilbert space, ${\cal  H}$.\\
{\bf A2.} The theory is invariant under inhomogeneous Lorentz transformations.\\
{\bf A3.} The energy-momentum of the states are defined. It follows from the
requirements of  Lorentz  and translation invariance that
we can construct a representation of the
orthochronous  Lorentz group. The representation
corresponds to unitary operators, $U( a, \Lambda)$,  and the theory is
invariant 
under these transformations. Thus there are Hermitian operators corresponding
to spacetime translations, denoted as $ P_{\mu}$, with ${\mu}=0,1,2,3,$ which have following
properties:
\be
\label{sec2.5}
\bigg[ P_{\mu},  P_{\nu} \bigg]=0
\ee
If ${\cal F}( x)$ is any Heisenberg operator then its commutator with $ P_{\mu}$
is
\be
\label{sec2.6}
\bigg[ P_{\mu}, {\cal F}( x) \bigg]=i\partial_{\mu} {\cal F}( x)
\ee
It is assumed that the operator does not explicitly depend on spacetime 
coordinates.
  If we choose a representation where the translation operators, $ P_{\mu}$,
are diagonal and the basis vectors $ |p,\alpha>$  span the Hilbert 
space,
${\cal  H} $, 
\be
\label{sec2.7}
 P_{\mu}| p,\alpha>= p_{\mu}| p,\alpha>
\ee
then we are in a position to make more precise statements: \\
${\bullet}$ Existence of the vacuum: there is a unique invariant vacuum state
$|0>$ which has the property
\be
\label{sec2.8}
 U( a, \Lambda)|0>=|0>
\ee
The vacuum is unique and is Poincar\'e invariant.\\
${\bullet}$ The eigenvalue of $ P_{\mu}$, ${ p}_{\mu}$,  
is light-like, with ${ p}_0>0$.
We are concerned  only with  massive stated in this discussion. If we implement
infinitesimal Poincar\'e transformation on the vacuum state then
\be
\label{sec2.9}
{ P}_{\mu}|0>=0,~~~ {\rm and}~~~ { M}_{\mu \nu}|0>=0
\ee
from above postulates and note that ${ M}_{\mu\nu}$ are the generators of Lorentz
transformations.\\
{\bf A4.} The locality of theory implies that a (bosonic) local operator 
at spacetime point
${ x}^{\mu}$ commutes with another (bosonic) 
local operator at ${ x}'^{\mu}$ when  their
separation is spacelike i.e. if $({ x}-{ x}')^2<0$. Our Minkowski metric convention
is as follows: the inner product of two 4-vectors is given by
${ x}.{ y}={ x}^0{ y}^0-{ x}^1{ y}^1-...-{ x}^{3}{ y}^{3}$.
Note that Hermitian conjugate of $\phi_n$, ${\phi_n}^{\dagger}=\phi_{-n}$.
By definition it  transforms as a scalar under inhomogeneous Lorentz
transformations
\label{sec2.10} 
\be  U( a, \Lambda)\phi_n( x) U( a, \Lambda)^{-1}=\phi_n(\Lambda  x+ a)
\ee
The micro causality, for any two local field operators,  $\Phi(x) $ and $\Phi(x')$ is stated to be
\label{sec2.11} 
\be
\bigg[\Phi( x),\Phi( x') \bigg]=0,~~~~~for~~( x- x')^2<0
\ee
{\it Remark:} Note that, for the compactified theory, the full Hilbert space is direct sum of Hilbert spaces, ${\cal H}_n$, designated by KK charges $n$.
Therefore, while deriving certain spectral representations, we have to sum over all states of the full Hilbert space subject to conservation laws. 

We define the amplitude as follows.The 
  incoming  particle state is $ |p_1,p_2 ~in>$ and outgoing state, also of two particles is $|k_2,k_1~out>$. Note that we are considering
  the inelastic amplitude. We mention in passing that the LSZ reduction 
procedure, spectral representation for retarded, advanced and
  causal commutators and their spectral representations hold good for 
elastic amplitudes as well as inelastic counterparts.
  We shall point out when there are distinctions between their 
analyticity properties at appropriate juncture.
  Let us consider the case when the two outgoing particle states are reduced.
  \bea
  \label{sec2.11}
 &&<k_2k_1, out|p_2p_1, in>-<k_2k_1,in|p_2p_1, in>=-{{i}\over{2\pi^4}}
 \int d^4x d^4y e^{+ik_1.x+ik_2.y}\nonumber\\&&
 (\Box_x+m_c^2)(\Box_y+m_c^2)
  <0|R\phi_n(x)\phi_{-n}(x)|p_2p_1> 
  \eea
 the retarded product is defined to be $R\phi_n(x)\phi_{-n}(y)= -\theta(x_0-y_0)[\phi_n(x), \phi_{-n}(y)]$. The matrix element of R-products are Lorentz
 and translation invariant.  Consequently, the latter invariance implies that it depend on difference of coordinates. We shall use this property
extensively. We define the scattering amplitude as
\bea
\label{sec2.13}c
<k_2k_1,out|p_2p_1,in>-<k_1k_2,in|p_2p_1,in>=2\pi i \delta^4(p_1+p_2-k_1-k_2)F(p_1,p_2,k_1,k_2)
\eea
Therefore,
\bea
\label{sec2.14}
F(p_1,p_2,k_1,k_2)=&&-\int d^4xd^4ye^{i(k_1.x+k_2.y)}(\Box_x+m_c^2)(\Box_y+m_c^2)<0|R\phi_n(x)\phi_{-n}(y)|p_2p_1,in>\nonumber\\ &&
=-\int d^4xd^4ye^{i(k_1.x+k_2.y)}<0|RJ_n(x)J_{-n}(y)|p_2p_1>
\eea
We have used equations of motion i.e. $(\Box_x+m_c^2)\phi_n(x)=J_n(x)$ 
where $J_n(x)$ is the source current. Moreover, we used a relation
$(\Box_x+m_c^2)(\Box_y+m_c^2)R\phi_n(x)\phi_{-n}(y)=RJ_n(x)J_{-n}(y)$. 
This relation, deserves following qualifying remarks since  
  additional terms are present on the right hand side which have not
been displayed explicitly. 
Notice that the two
Klein-Gordon operators $ (\Box_y+m_c^2) (\Box_y+m_c^2)$ act on the 
R-product;  $R\phi_n(x)\phi_{-n}(y)$. It is evident that the
term  $RJ_n(x)J_{-n}(y)$ would appear. 
In addition, the operation of KG operators on $R$-product
of two field operators  will give  derivatives of $\delta$-functions.  
Moreover, there  
will be equal time commutators of operators. 
Notice that derivatives of delta function when Fourier transformed will 
give products of momenta.  However,
these products have to be Lorentz invariant since $S$-matrix elements are Lorentz invariant. Therefore,  such terms can be expressed in terms of 
Mandelstam variables, $s,t$ and $u$. As far as
study of analyticity properties of the amplitude is concerned, presence of 
such terms (polynomials in s,t and u)  do not affect us; moreover, 
for local field theories,
only finite number of such $\delta$-function derivative terms will appear. 
 The third type of terms are equal time commutator of operators. 
These terms vanish from microcausality arguments: 
 $[{\cal O}(x_0, {\bf x}),{\cal O}(y_0,{\bf y})]\delta(x_0-y_0)=0$.  

We have denoted the fields arising from reduction of out states as $\phi_n(x)$ and $\phi_{-n}(x)$; one carries KK charge of $n$ units and other carries  charge
of $-n$ units; recall the momentum along compact direction, 'charge' is quantized as $n\over R$ where $R$ is radius of the circle.  We may reduce one particle
from 'out' state and another from 'in' state. In order to study analyticity property of amplitudes with requirements of Lorentz invariance and microcausality, it
suffices to consider reductions of two states in whatever combinations we desire. We shall focus attention on the LSZ reduced amplitude (\ref{sec2.11}) in what
follows. Moreover, we could reduce all the four particle states i.e. $|k_2k_1,out>$ and $|p_2p_1,in>$; however, as is well known, in order to study analyticity
properties and to write down dispersion relations, it is most convenient to reduce two particles from 'in' or 'out' states. Let us define three distributions as
follows. We shall recognize shortly the importance of the three distributions.
\bea
\label{sec2.15}
F_R(q)=\int d^4\xi e^{iq.\xi}\theta(\xi_0)<P_f| [J_n(\xi/2),J_{-n}(-\xi/2)]|P_i> \nonumber\\
F_A(q)=-\int d^4\xi e^{iq.\xi}\theta(-\xi_0)<P_f| J_n(\xi/2),J_{-n}(-\xi/2)] |P_i> \nonumber \\
F_C(q)=\int d^4\xi e^{iq.\xi}<P_f|[J_n(\xi/2),J_{-n}(-\xi/2)]|P_i>
\eea
Here $|P_i>$ and $|P_f>$ are arbitrary momentum states. We may treat $P_i$ and $P_f$ as parameters for our purpose since we do not intend 
to reduce them.  There are several pertinent points which 
need mentioning. First thing to note is that each of the distributions 
when Fourier transformed
respectively to 
$\tilde{F}_R(\xi), {\tilde F}_A(\xi) ~{\rm and}~{\tilde F}_C(\xi)$ vanish for 
$\xi^2<0$.

We briefly discuss a few aspects of domain of analyticity of the distributions, defined above, as a function of complex $q$. It will suffice to focus
attention on $F_R(q)$ since it is related to the amplitude we introduced earlier. Moreover, these arguments are easily extended to study of analyticity
of $F_A(q)$ and $F_C(q)$.
  Notice that the Fourier transform  $F_R(q)$, ${\tilde F}_R(\xi)$,
 is defined in the forward light cone, $V^+$, i.e. $\xi_0>0$ and $\xi^2>0$ and it  vanishes for $\xi^2<0$ due to microcausality. Let us turn attention to properties of the function, $F_R(q)$ for complex $q$. The function will be analytic in the complex plane if following properties are satisfied. (i) The exponential must converge in
 all directions in complex plane for $\xi\in V^+$. Thus  we desire that $Im~q>0$ so that exponential is damped in the upper half complex $q$-plane; 
 however, there is no restriction on real part of $q$.  Moreover,
 for $\xi\in V^+$, it is necessary that $Im~q\in V^+$ also i.e. $(Im~q)^2>0~ {\rm and}~ Im~q_0>0$. This defines a forward tube $T^+$. Similar arguments
 is used to define the domain of holomorphy of $F_A(q)$ whose domain of holomorphicity, in the complex $q$-plane, is defined in the backward light cone,
 $V^-$, and the corresponding tube is $T^-$. We shall argue in what follows that both $F_R(q)$ and $F_A(q)$ vanish for unphysical real values of $q$-variables
 and,  here $F_R(q)=F_A(q)$ since $F_C(q)=0$. Let us open up the commutators of the two currents in the expression for $F_C(q)$. Let the first term
 of the two be $F_{1C}(q)$ which is expressed as
 \bea
\label{sec2.16}  
F_{1C}(q)=\int d^4\xi e^{iq.\xi}<P_f|J_n(\xi/2)J_{-n}(-\xi/2)|P_i>
\eea
Now we insert a complete set of physical states, 
$\sum_{p_N,q_N} |p_N,q_N><p_N,q_N|={\bf 1}$, where $q_N$ 
refer to all other quantum numbers 
designating the state
\bea
\label{sec2.17}
F_{1C}(q)=\int d^4\xi e^{iq.\xi}\sum_{N,q_N}<P_f|J_n(\xi/2)|N,q_N>
<N,q_N|J_{-n}(-\xi/2)|P_i>
\eea 
Note that currents carry KK charges. Therefore, the matrix elements such as 
$<P_f|J_n(\xi/2)|N,q_N>$ and $<N,q_N|J_{-n}(-\xi/2|P_f>$ will be nonvanishing
only when all conservation laws are respected. Next we go step by step 
as follows: (i) use translational symmetry transformations on
each of the product matrix elements of (\ref{sec2.17})  such that 
$J_n(\xi/2)\rightarrow J_n(0)$ and  
$J_{-n}(-\xi/2)\rightarrow J_{-n}(0)$
to bring the  argument of each current   to zero. 
(ii) Recall that translation operator $e^{-iP.a}$ acting on a state of momentum states, $|p>$ brings out a factor $e^{-ip.a}$.
Thus there will be four such factors coming from this operation and an additional factor of $e^{iq.\xi}$. (iii) When we integrate over $\xi$, there is a 
$\delta$ function which implies that $p_N=1/2(P_f+P_i)-q$.  The same arguments go through for the second matrix element which appears
after the commutator is opened up; the only difference is that product of currents are in reverse order: $J_{-n}(-\xi/2)J_n(\xi/2)$. Now the delta function arising
out of $\xi$ integration leads to $p_N=1/2(P_i+P_f)+q$. We express (\ref{sec2.15}) as
\bea
\label{sec2.18}
F_C(q)=&&\sum_{N}\bigg[<P_f|J_n(0)|p_N={{(P_i+P_f)}\over 2}-q,q_N>
<p_N={{(P_i+P_f)}\over 2}-q,q_N|J_{-n}(0)|P_i>\nonumber\\&& 
-<P_f|J_{-n}(0)|p_N={{(P_i+P_f)}\over 2}+q,q_N><p_N={{(P_i+P_f)}\over 2}+q,q_N|J_{n}(0)|P_i>\bigg]\nonumber\\&&
\eea
It is important to emphasize that the intermediate complete set of states, 
$|p_N, q_N>$ are physical and satisfy mass-shell conditions, 
${{P_i+P_f}\over 2} \in V^+$ where $V^+$ is the forward 
lightcone.
Therefore, it follows that there are states for each of the terms 
on the $r.h.s$ of (\ref{sec2.18}) which satisfy the following conditions
\bea
\label{sec2.19}
({{P_i+P_f}\over2}\pm q)^2\ge {\bf M_{\pm}}^2,~~{\rm and}~~({{P_i+P_f}\over 2}\pm q)_0\ge 0
\eea
In other words, for each of the matrix  elements there exists a minimum mass state with positive energy. If this condition is not satisfied
by one of the matrix elements that is  $({{P_i+P_f}\over 2} + q)^2< {\bf M_{+}}^2,$ and $({{P_i+P_f}\over 2}+ q)_0< 0$
then it vanishes. 
Similarly, when 
$({{P_i+P_f}\over2} - q)^2< {\bf M_{-}}^2,$ and $({{P_i+P_f}\over 2}-q)_0< 0$ 
then it also vanishes. Therefore, there
are two regions for the two matrix elements where each  can vanish. 
There is a third region where the above conditions (\ref{sec2.19}) are
not satisfied for each of the matrix elements and consequently,  
both matrix elements vanish simultaneously. This is unphysical region for both. The range of $q_0$ is
easily determined as follows.  We go to a frame where 
$1/2(P_i+P_f)=(m_0, {\bf 0})$. 
Thus $[1/2(P_i+P_f)\pm q]^2=(m_0 \pm q_0)^2\pm ({\bf q})^2$.  
Thus the region where both matrix elements vanish 
we have
 a relation: 
$(m_0-\sqrt{{\bf q}^2+m_1^2})\le q_0\le(-m_0+\sqrt{{\bf q}^2+m_2^2})$ for 
real values of momenta.
 
The important question to ask is what is the singularity free region 
in complex $q$ plane? Jost and Lehmann \cite{jl} and, independently, 
Dyson \cite{fd} (JLD)
obtained a representation for $F_C(q)$ as well as, by extension, for $F_R(q)$ 
which enables us to find a singularity free  domain in complex 
$q$-plane\footnote{See Schweber \cite{ss} for elaborate discussions on 
JLD theorem.}. Lehmann \cite{lehmann} proved existence of two ellipses, 
known as small Lehman ellipse (SLE) and large Lehmann ellipse (LLE). The 
scattering amplitude
 converges for complex $t$ inside SLE. On the
 other hand  absorptive part of the amplitude  converges inside larger
 ellipse (LLE) defined in the complex $t$-plane.
In other words, the partial wave expansion has a larger domain of converges than
 $-1\le cos\theta\le +1$.  We present below definitions of Lehmann ellipses\\
 (i) SLE
 \bea
 \label{sec2.20}
 E_{SLE}^{ab\rightarrow cd}:~~: | t + (P_1 - K_1)^2+|t + (P_1+K_1)^2| < 4|{\bf P}_1|{\bf K}_1| x_0(s)
 \eea
 where $m_a=m_b$ and $m_c=m_d$, for the case at hand. Furhermore,
 \bea
 \label{sec2.21}
 x_0=\bigg[1+{{({\bf M}_a^2-m_a^2)({\bf M}_c^2-m_c)^2}\over{K_1(s)^2s}}\bigg]^{1/2}
 \eea
 We have used ${\bf M}_a={\bf M}_b$ and ${\bf M_c}={\bf M}_d$ in 
arriving at the above formula. Note that $x_0>1$. \\
 (ii) LLE
 \bea
 \label{sec2.22}
 E_{LLE}^{ab\rightarrow cd}: | t + (P_1 - K_1)^2+|t + (P_1+K_1)^2| <4 {\cal Z}_L(s)
 \eea
 where ${\cal Z}_L(s)$ is given by
 \bea
 \label{sec2.23}
 {\cal Z}_L(s)=&&\bigg(\bigg[{{({\bf M}^2_a-m_a^2)^2}\over{s}}\bigg]
 \times\bigg(\bigg[{{({\bf M}_c^2-m_c^2)^2}\over{s}}\bigg]\bigg)^{1/2}\nonumber\\&&
  + \bigg(\bigg[{{({\bf M}^2_a-m_a^2)^2}\over{s}}+P_1(s)^2\bigg]^{1/2}\times
\bigg[{{({\bf M}_c^2-m_c^2)^2}\over{s}}+K_1(s)\bigg]^{1/2}\bigg)
 \eea
 We note that LLE is bigger than SLE.  The absorptive part of the 
 scattering amplitude, as noted, converges inside LLE. 
 In fact this proof of the
 existence of LLE by Lehman was very crucial to write down fixed-$t$ dispersion relations. It is to be noted, as we shall see when we
 write dispersion relations, that the integral is over values of $s$
 from $s_{th} ~to~ \infty$. Therefore, there is a LLE for every $s$. 
 Moreover, the ellipse
 shrinks to a real line segment of $t$-plane in the limit $s\rightarrow\infty$. 
  All the
 results summarized above follow by demanding Lorentz invariance and microcausality and other LSZ axioms except unitarity.
 Martin \cite{book2,martin66} invoked unitarity and utilized the positivity properties of partial wave amplitudes of elastic  scattering to derive stronger results.
 The study of analyticity in the framework, mentioned above to derive
results (i) and (ii),
 is called linear program (procedure) since unitarity of S-matrix is not invoked until now. 
 Martin's work was influenced by a paper of Lehmann \cite{lehmann66}. 
The discussions presented so far, in a concise form, were derived for a 
scalar field theory
 which originated   as follows.
 The original theory, to remind the reader, was  considered in $D=5$ which was compactified to $R^{3,1}\otimes S^1$. We have considered
 4-dimensional fields $\{\phi_n \}$ to derive above results.  
 
  We need
 one more ingredient for writing dispersion relations as is established at 
present. The dispersion relations are written down by closing a contour
 in the complex $s$-plane. Therefore, it is necessary to know what kind 
of singularities lie in this plane and their locations. This questions was
 addresses and answered by Bross, Epstein and Glaser \cite{beg}. 
They considered a two body scattering amplitude,
  $a+b\rightarrow c+d$. They demonstrated that the (possible) complex 
singularities are isolated and  located in finite region of $s$-plane
(usually called a'potato').  
They argued
  that there will exist a path on   which scattering amplitude 
for $a+b\rightarrow c+d$ can be continued analytically to the u-channel
  physical scattering amlitude i.e. $a+{\bar d}\rightarrow c+{\bar b}$.  
They also noted that the size of the domain is not fixed;
  it depends on $t$, however.  They estimated the growth property of this domain as $t$ grows. The growth is $|t|^{3+\epsilon}, \epsilon>0$ and
  the parameter is arbitrary. We mention that BEG theorem is valid for 
elastic as well as for inelastic processes and will be used as
an useful ingredient for us.
  Now we   shall state Martin's theorem. 
The fixed-t dispersion relation, without subtraction, is
  \bea
  \label{sec2.24}
  F(s,t)={1\over{\pi}}\int_{s_{thr}}^{\infty} ds'{{Im F(s',t})\over{s'-s}}+
  {1\over{\pi}}\int_{{u_{thr}}}^{\infty} du'{{Im F(u',t)}\over{u'-u}}
  \eea
  when $t\in$LLE.
  
  \bigskip
  
  \noindent{\it Martin's Theorem} \cite{martin66} \\
  
 \noindent  If the elastic amplitude (consider equal mass scattering of mass, m,  for simplicity) satisfies following conditions:\\
  1. $F(s,t)$ satisfies fixed-t dispersion relations  in $s$ with finite number of subtractions, $N\le 2$ with   $-t_M\le t\le 0$ where 
  \bea
  \label{sec2.24}
  t_M={{4\over{s'>min({\bf{M}_{ab}^2,{\bf M}_{ac}^2})}}\bigg[P_1(s')^2+{{({\bf M}_a^2-m_a^2)^2}\over{s'}}}\bigg]
  \eea
  Note that $t_M>0$.\\
  2. $F(s,t)$ is an analytic function of $s$ and $t$ (in both the variables) in the neighborhood of any $\bar s$ in some interval below the
  physical threshold: $s_{thr}-\eta<4m^2$  and in some neighborhood $|t| <R({\bar s})$ of $t=0$ \cite{beg,lehmann66}. \\
    3. $A_s(s',t)$ and $A_u(u',t)$, the absorptive parts of $F(s,t)$ from discontinuities of right  hand cut, $s'>4m^2$ and left hand cut, $u'>4m^2$,
  satisfy following positivity properties. They are also holomorphic in LLE. \\
  4. The positivity properties of $A_s(s',t)$ and $A_u(u',t)$ for $s'>4m^2$ and $u'>4m^2$ are
  \bea
  \label{sec2.25}
  |({{\partial}\over{\partial t}})^nA_s(s',t)|\le ({{\partial}\over{\partial t}})^nA_s(s',t)|_{t=0},~~~-4k^2\le t\le 0
  \eea
  and
  \bea
  \label{sec2.26}
   |({{\partial}\over{\partial t}})^nA_u(u',t)|\le ({{\partial}\over{\partial t}})^nA_u(u',t)|_{t=0},~~~-4k^2\le t\le 0
\eea
Martin proved, based under assumptions (1)-(4) that $F(s,t)$ is analytic in a domain of quasi topological product
\bea
\label{sec2.27}
\bigg\{s,t \bigg| s\in{\rm cut plane}, s\ne(4m^2+\rho, \rho>0~~{\rm and}~~ s\ne 0-t-\rho', \rho' >0\\
|t|<2R~~{\rm with ~some~R}  \bigg\}
\eea
The consequence of Martin's theorem is that there exists a radius, $R_M$, such that the dispersion relation is valid for $|t|<R_M$. The domain
of analyticity of $F(s,t)$ is the quasi-topological product, s-t cut-plane$\otimes$ a circle in t-plane with center at $t=0$  and the radius is $|t|=R_M$.\\
{\it Determination of $R_M$}:  Martin used analyticity arguments on the 
amplitude to determine $R_M$ and proved that it is
independent of $s$ for large $s$, 
\bea
\label{sec2.28}
R_M={\rm min} [t_M,{max}~{{r_{sL}(s')}\over{s'}}]
\eea
where $r_{sL}(s')=2P_1(s')^2(x_0(s')-1)$. 
Here ${\bf P}_1(s')$ is the c.m. momentum and its magnitude is the same for both incoming pair and outgoing pair for the case at hand. 
$r_{sL}$ corresponds to the right extremity of the large Lehmann ellipse.  
We shall discuss the analyticity domain in the $t$-plane for inelastic scattering
where Martin's theorem cannot be adopted directly since the absorptive inelastic amplitudes do not satisfy the positivity conditions (\ref{sec2.25}) and
(\ref{sec2.26}).  Therefore, a chain of arguments will be presented to study analyticity domains in $s$ and in $t$ planes.

\bigskip

{\section{ Analyticity Properties of Inelastic Scattering Amplitudes}} 

\bigskip

We study the analyticity properties of inelastic process $a+b\rightarrow c+d$ in this section. It is assumed that $a$ and $b$ are equal mass particles,
$m_a=m_b$ and so are the two final state particle, $m_c=m_d$. Moreover, it is also assumed that produced particles are heavier than accelerated
initial colliding  particles i.e. $m_c>m_a$. 
A point to note is that   $s_{thr}\ne s_{phys}$. 
Consequently, the analyticity properties of inelastic reactions are
to be analyzed carefully and the results derived for elastic scattering do not automatically hold good here.  A few additional steps are necessary to derive
various characteristics of the inelastic amplitude, $F^{ab\rightarrow cd}$. The first important point in this case, which differs from elastic case,  is that
the inelastic partial wave amplitudes, $f_l(s)^{ab\rightarrow cd}$ do not enjoy the positivity properties of elastic partial wave amplitude i.e.
$0\le |f_l(s)^{ab\rightarrow ab}|^2\le Im~f_l(s)^{ab\rightarrow ab}\le 1$.  We also know that there is no analog of optical theorem for inelastic
amplitudes.  We recall that analyticity properties of inelastic amplitudes have been studied,  in the past , in a rigorous framework  
\cite{russian1,russian2}. These authors assumed that ${\rm Im}~F(s,t)$
is analytic inside an ellipse demanding certain inequality, 
to be satisfied by partial wave amplitudes. They had not determined the 
properties of their ellipse from basic principles.
However, once the existence of Lehmann-Martin
ellipse is proved, as $s\rightarrow\infty$, in the complex $t$-plane there 
is a domain which does not shrink to real line segment as was proved 
by Martin
\cite{martin66}. Indeed, the proof of existence of Martin's radius, $R_M$, 
is crucial in the sense that $R_M$ is independent of $s$.  In view of
preceding remarks, the results derived by those authors 
\cite{russian1,russian2}, at that juncture, were based not on strictly 
rigorous arguments.    
Subsequently, Sommer \cite{sommer67} carefully studied analyticity property 
in inelastic reaction for hadronic production processes. He obtained
$t$-plane domain of analyticity with refinement of Martin's work 
\cite{martin66}. We shall adopt similar line of arguments to 
study production of KK states
in high energy collisions at LHC. Notice, however, the difference, 
Martin \cite{martin66} was able to determine  value of the radius of 
the circle, $R_M$, 
from first principles
in terms of the experimentally observed mass. 
For most of the cases, it was determined to be $t_0=4m_{\pi}^2$.  
For the case under study we cannot 
identify the analog of $R_M$ to an experimentally measured number and we 
can only present it in terms of the mass parameters of the theory.  
From our perspective,
we determine the domain of analyticity and we write fixed-t dispersion relations for  the right hand cut $F(s,t)^{ab\rightarrow cd}$.  One more step is to be completed.
We must determine the analyticity property of the $u$-channel amplitude, $F(u,t)^{a{\bar d}\rightarrow c+{\bar b}}$, to write the dispersion relation 
for the left hand cut.  Therefore, it is necessary to go through the same procedure, as is adopted for the s-channel reaction, in order to determine the
domain of holomorphy in $t$. 

We have discussed the existence of Martin's ellipse for elastic amplitude  
and the domain of convergence  of the absorptive amplitude ; the
semi-major axis of the ellipse is $cos\theta_0=1+{{t_o}\over{2P_1(s)^2}}$. 
We have summarized the results in the preceding section.   
It follows from unitarity arguments that 
$\sigma_t(s)^{ab}=\sigma_{el}(s)^{ab} +\sigma(s)^{ab\rightarrow cd}+
{\rm all ~ inelastic~channels}$.
Also note that ${\rm Im}~f_l(s)^{ab\rightarrow ab}
=|f_l(s)^{ab\rightarrow ab}|^2+|f_l(s)^{ab\rightarrow cd}|^2+....$ where 
$(...)$ stand for other two body
amplitudes. This is due to the fact ${\rm Im}f_l(s)^{elastic}$ is 
related to total cross section which is sum of elastic and all inelastic cross 
sections (when we write in terms of partial wave amplitudes). 
Thus we write inequality for imaginary part of inelastic partial wave amplitude as
\bea
\label{sec3.1}
 |f_l(s)^{ab\rightarrow cd}|\le {\sqrt{{\rm Im}~f_l(s)^{ab\rightarrow ab}}}
 \eea
since partial wave amplitudes $f_l(s)^{ab\rightarrow cd}$ do not enjoy positivity properties anymore.  We also know that, due to polynomial boundedness
of scattering amplitude and convergence of partial wave expansion inside Martin ellipse, the elastic partial wave amplitude has following behavior for asymptotic energy as $l$ tends
to large values. Note that $P_l(1+{{t_0}\over{2K_1(s)^2}})$ grows exponentially since $cos\theta>1$ in this region. Therefore, the imaginary part of partial
wave amplitude must be damped for large $l$ when we take $s$ to be large
\bea
\label{sec3.2}
 lim_{l\rightarrow\infty}{\rm Im}~f_l(s)^{ab\rightarrow ab}\le \bigg[{{1\over{1+{R_M\over{2P_1(s)^2}}+{\sqrt{(1+{{R_M}\over{2P_1(s)^2}}-1}}}}} \bigg]^l
 \eea
 The inelastic amplitude $F(s,t)^{ab\rightarrow cd}$ can be expanded in partial wave expansion, taking into account the inequality satisfied by imaginary
 part of its partial wave (\ref{sec3.1}). The partial wave expansion is 
 \bea
 \label{sec3.3}
 F(s,t)^{ab\rightarrow cd}(s,t)={{\sqrt{s}}\over{|{\bf P}_1(s)|}}\sum_{l=0}^{\infty}(2l+1)f_l(s)^{ab\rightarrow cd}P_l(cos\theta)
 \eea
 Notice that the prefactor ${{\sqrt s}\over{|{\bf P}_1(s)|}} \rightarrow 1 ~{\rm as}~s\rightarrow\infty$. We shall not generally include such constant prefactors
 at various places unless explicit mention of them carries important implications.
 We recall that for the inelastic scattering $cos\theta={{t+P_1(s)^2+K_1(s)^2}\over{2P_1(s)K_1(s)}}$. The argument presented above is similar
 to the arguments of \cite{russian1,russian2}; however,  the power of Martin's theorem and the existence of Martin'e ellipse  was not recognized by them.
 Moreover, one of their primary goal was to rigorously study properties of inclusive processes.
 In fact it is not necessary to assume existence of an ellipse inside which inelastic partial wave converges.  Indeed, with preceding arguments, it can be proved 
 that $F(s,t)^{ab\rightarrow cd}$ is analytic inside an ellipse which is defined, in the $t$-plane by the condition
 \bea
 \label{sec3.4}
 |t+({\bf P}_1(s)-{\bf K}_1(s))^2|+|t+({\bf P}_1(s)+{\bf K}_1(s))^2|<4|{\bf P}_1(s)||{\bf K}_1(s)|{\sqrt{{1+{{R_M}\over{4P_1(s)^2}}}}}
 \eea
 We mention in passing that, for elastic process, ${\bf P}_1(s)={\bf K}_1(s)$  and thus $(P_1(s)-K_1(s))^2=0$.
 Notice that as $s\rightarrow\infty$, the right hand side is ${R_M}\over 4$ and is independent of $s$.  We have argued in Section 2 that 
 ${\rm Im}~ F(s,t)^{ab\rightarrow cd}$ is analytic inside LLE besides being analytic in Martin's ellipse.  In the high energy limit, we can see 
that Martin's ellipse is bigger than LLE; however, in the low $s$ region, the case is opposite.  We noted earlier that the domains defines by SLE and
LLE depend on $s$.  Thus, while writing dispersion relations (integral over $s'$-variable) and  identifying domain of holomorphy in $t$,  this aspect
should be kept in mind in the sense that the domain depends on $s'$. Since we are considering unequal mass scattering the integration 
regions of $s'$ variable is to be taken into considerations. (i) $s_{thr}\le s'\le s_{phys}$, (ii) $s'\ge s_{phys}$. (iii) Another region is to be considered
due to the complications of unequal mass inelastic scattering i.e. masses of initial colliding (equal mass) particles producing particle 
(although they are of equal mass) whose masses are
different from initial colliding particles (that is why $s_{thr}$ and $s_{phys}$ are not equal).  This region is called $s$-critical region which has been
discussed in detail by Sommer \cite{sommer67}. For our problem this value is determine from the relation
\bea
\label{sec3.5}
|{\bf K}_1(s)|{\sqrt{{\bf P}_1(s)^2+{R_M\over 4}}}=\zeta(s_{sc})
\eea
where
\bea
\label{sec.5}
\zeta(s)=&&\bigg[{{({\bf M}_a^2-m_a^2)^2}\over s}\bigg]^{1/2}\times\bigg[   {{({\bf M}_c^2-m_c^2)^2}\over s}\bigg]^{1/2}+\nonumber\\&&
[{{({\bf M}_a^2-m_a^2)^2}\over s}+P_1(s)^2\bigg]^{1/2}  \times \bigg[   
{{({\bf M}_c^2-m_c^2)^2}\over s}+K_1(s)^2\bigg]^{1/2}
\eea
A remark is in order at this stage. In case of hadronic reactions, 
${\bf M}_a, {\bf M}_c, m_a$ and $m_c$ are obtained from experimental data
from data tables. Therefore, one can use numerical factors to determine the critical value.  We do not have values of these mass parameters 
(except we may choose that $m_a$ is mass of proton).  However, as 
we shall discuss in the sequel, 
the growth of cross sections as a function of $s$ will be
bounded. We mention that for $s'>s_{cr}$ LLE is inside Martin's ellipse.  
Now, for region  (i), $s_{thr}\le s' \le s_{phys}$, 
the domain of holomorphy is LLE.
We can write fixed-$t$ dispersion relation for $-t_M\le t\le -t_m$ (see the definition of $t_m$ below). 
We denote this domain as ${\cal D}_1$ where 
\bea
\label{sec3.6}
{\cal D}_1={\large\cap}_{s_{thr}\le s'\le s_{phys}}E_L(s')^{inel}
\eea
The second domain of holomorphy, ${\cal D}_2$, is where the Lehmann ellipse is inside the Martin's ellipse. Therefore, in the
$s\rightarrow\infty$ limit this complex-$t$ domain does not degenerate to a real line
\bea
\label{sec3.7}
{\cal D}_2={\Large\cap}_{{s_{cr}}\le s'<\infty} E_{Martin}(s')^{inel}
\eea
This is defined when $t$ lies in the interval $-t_M\le t\le t_m $(see more discussions below). Moreover, in this $s'$ region, each Martin's ellipse,
$E_{mart}(s')$ is larger than corresponding LLE, $E_L(s')$. Furthermore, for asymptotic $s'$,  $R_M(s)\rightarrow R_M$ where $R_M$ is
$s$-independent. As asserted above, non of the Martin's ellipses collapse to real $t$-axis and we conclude the ${\cal D}_2$ is a complex domain. Let us
look at the interval $s_{phys}<s'\le s_{cr} $ and denote the corresponding domain as ${\cal D}_3$. It is given by
\bea
\label{sec3.8}
{{\cal{D}}_3}={\Large{\cap}}_{{s_{phys}}<s'\le s_{cr}} E_L^{inel}
\eea
Here $-t_M\le t\le -t_m$. Notice that $s'$ is restricted to a finite interval. Consequently, the complex domain in $t$-plane is not permitted to shrink
real $t$ line.  We note that it is necessary to find the two extreme values of real $t$ within which interval $t$ lies; 
we  remit $t$ to go out of physically
allowed values. However, it is restricted to above interval. The delicate issues arising for unequal mass inelastic scatterings 
is discussed briefly
in what follows.
\bea
\label{sec3.9}
t_m=&&max\bigg\{max_{s\ge{\bf M_{ab}}^2}\bigg[ 2P_1(s)^2+2({{m_a^2-m_c^2}\over{4s}})-2{\cal Z}_L(s)\bigg], \nonumber\\&&
max_{u\ge {\bf M}_{ac}^2}\bigg[{\bar P}_1(u)^2+
{ \bar K}_1(u)^2+2({{m_a^2-m_c^2}\over{4u}})-2{\cal Z}_L(u) \bigg]\bigg\}
\eea 
where ${\cal Z}_L(u)$ is an expression similar to ${\cal Z}(s)$ (\ref{sec2.23}) and directly related to the $u$-channel process.
\bea
\label{sec3.10}
{\cal Z}_L(u)=&&\bigg[{{({\bf M}_a^2-m_a^2)({\bf M}_c^2-m_c^2)}\over{u-({\bf M}_a^2-{\bf M}_c^2)}}\bigg]+
\bigg[{{({\bf M}_a^2-m_a^2)({\bf M}_c^2-m_c^2)}\over{u-({\bf M}_a^2-{\bf M}_c^2)}}+{\tilde P}_1(u)^2\bigg]^{1/2}\times\nonumber\\&&
\bigg[{{({\bf M}_a^2-m_a^2)({\bf M}_c^2-m_c^2)}\over{u-({\bf M}_a^2-{\bf M}_c^2)}}+{\tilde K}_1(u)^2\bigg]^{1/2}
\eea
where ${\tilde P}_1(u)$ and $\tilde{K}_1(u)$ are the c.m momenta of initial pair of states and final pair of states in the $u$-channel reaction. These formulae,
(\ref{sec3.6}) and (\ref{sec3.7}) arise in case of unequal mass inelastic scattering and the corresponding formulae for elastic case assume rather
simple form.  A few important comments deserve mention at this juncture. (i) One has to be careful to check that the c.m. momenta for direct channel
and c.m. momenta for crossed channel process do not become complex. 
In such a situation there will be a gap in $s$ where we are unable to
write dispersion relations. Same argument is valid for the cross channel. (ii) A situation might arise  when ${\cal Z}_L(s)$ might become complex and/or
${\cal Z}_L(u)$ might be complex in some region. 
In such a situation the Lehmann would not be defined (i.e. will not exist) 
in those intervals of $s$ and/or $u$.  In absence of Lehmann ellipse for 
some interval of $s$ and/or $u$ we cannot write fixed-$t$ dispersion relations.  It is necessary that $t_M>t_m$. We record
these comments at this stage to draw attention of the reader to 
the delicate issues to be addressed, in the context of study 
of analyticity properties,
for   inelastic unequal mass scattering in general. We also note that, in the study of hadronic scattering, the masses of particles, 
involved in scattering,
are known. Therefore, the values obtained from experiments are utilized to chart the domains. For the case at hand, it is not possible to actually
display the forbidden domains. It is to be noted that, in the high energy limit, when $s$ and $u$ take asymptotic values, the formulae are 
considerably  simplified.  The purpose of this detour, in our discussions,
 is that in the works of Russian group 
\cite{russian1,russian2} these delicate
arguments were not presented. Therefore, there were, to some extent additional assumptions in their work; moreover, domains of analyticity were not charted out
carefully for unequal mass case involving inelastic reactions. Those aspects have been taken into account in analyzing  
domains of holomorphy with care
in the present investigation.  

We have identified the domains of analyticity in various values of $s'$ in order to write down dispersion relations. 
Our focus has been to identify the
holomorphic domain for writing a dispersion relation associated with the right hand cut, the $s$-channel reaction, $a+b\rightarrow c+d$; it is
\bea
\label{sec3.11}
{\cal I}_R={{1\over{\pi}}}\int_{s_{thr}}^{\infty}ds'{{{\rm Im}~F(s',t)^{ab\rightarrow cd}}\over{s'-s}}
\eea
 This is an unsubtracted dispersion relation. The scattering amplitude is polynomially bounded in $s$ for $t$ lying within Lehmann-Matrin ellipse i.e.
 $|F(s,t)^{ab\rightarrow cd}|\le s^N$ and $N\le 2$ as follows from Jin-Martin 
theorem \cite{jin-martin}. Therefore, if at all we need to write
a subtracted dispersion relation, it will need at most two
subtractions. Moreover the fixed-$t$  dispersion 
relation is valid in the domain of
 complex $t$-plane given by
 \bea
 \label{sec3.12}
 {\cal D}^R_{inel}=\bigg[\large\cap_{s_{thr}\le s'\le {s_{phys}}} E_L^{inel}(s') \bigg]{\Large\cap}_{s'\ge s_{phys}}\bigg[\cap \bigg(E_M^{inel}(R,s') \bigg)\cup E_L^{inel}(s') \bigg]
\eea
Here $R$ stands for right hand side of the threshold where we identify domain of holomorphicity in complex $t$-plane.
We have argued that the union of complex domains is such that each of them does not shrink to real $t$-line. In order to write dispersion 
relation for
$F(s,t)^{ab\rightarrow cd}$,  we have to consider properties of the $u$-channel amplitude. Therefore, we consider the reaction 
$a+{\bar c}\rightarrow d+{\bar b}$.
Thus scattering of two particles of mass $(m_a,m_c)$ to $(m_c,m_a)$ is 
to be considered, (note $m_a=m_b$ and $m_c=m_d$).  
This process is practically, as far as kinematics is concerned,
 like elastic scattering of unequal mass particles.  We recall that masses 
of anti particles $({\bar b}, {\bar d})$ respectively are 
$(m_a,m_c)$.  We can determine
the semi-major axis of the $u$-channel Martin ellipse and SLE and LLE in this case.  It is obvious that we do not encounter any technical
difficulties in determining the $t$-plane analyticity domain for the crossed channel reaction. Note that the magnitudes of c.m. momenta of initial pair and final
 pair are the same. The mathematical expressions are 
exactly the same as those for elastic process. Consequently, calculations are simplified to identify the domain of holomorphicity of $t$. The unsubtracted dispersion
relation for the inelastic amplitude is (including  contributions from right hand cut and left hand cut) is
\bea
\label{sec3.13}
F(s,t)^{ab\rightarrow cd}={1\over{\pi}}\int_{s_{thr}}^{\infty}ds'{{{\rm Im}~F(s,t)^{ab\rightarrow cd}}\over{s'-s}}+{1\over{\pi}}\int_{u_{thr}}^{\infty}du'
{{{Im }~F(u',t)^{a{\bar c}\rightarrow d{\bar b}}}\over{u'-u}}
\eea

\bigskip

\noindent{\bf Bounds on Cross Sections for Inelastic Reaction} 
${\bf F(s,t)^{ab\rightarrow cd}}$

\bigskip

We proceed to derive bounds on cross sections utilizing the analyticity properties of the amplitudes derived thus far. The first point to note is  that 
we are unable to derive analog of the Froissart bound for inelastic total cross sections $\sigma(s)_t^{inel}$ from basic principles. 
The reason is that
the absorptive amplitude, ${\rm Im}~F(s,t)^{ab\rightarrow cd}$ of two body reaction lacks the positivity properties of its elastic counter part. 
Thus the insight gained in the preceding discussions will be utilized to derive bounds of our interests. Furthermore, we note that 
$F(s,t)^{ab\rightarrow cd}$
is polynominally bounded, $s$, for fixed $t$.  Thus it is possible to derive upper bounds on differential cross sections. Logunov and collaborators
\cite{russian1,russian2} derived such upper bounds; however, an unknown constant appeared in such a bound. 
They did not employ  the full power of the analyticity properties of 
amplitude in $s$ and $t$ variables. 
Singh and Roy \cite{sr70}.    
used Lagrangian multiplies technique to derive very powerful bound. One of 
their important results was that the bound on elastic
differential cross section was obtained  without any unknown constant. 
Moreover, the bound is expressed in terms of experimentally measured 
quantities
such as $\sigma_t$ and $\sigma_t^{el}$. Indeed the bound on, differential  
elastic cross section, ${{d\sigma^{el}}\over{d\Omega}}$,
 was tested against experimental data over a wide range of high energy scattering experiments 
 and no violations were found.  Furthermore, all ingredients that went 
into deduce the bound have been proved, in the past, 
from axiomatic field theory.
In case of inelastic scattering not all those results hold good.  We shall 
utilize the rigorous results reported here and present them which will not be
as strong as  on the elastic cross sections
which is not a surprise. Moreover, one of our 
limitations is that, at present, we are unable to test the bounds 
against experimental
data.  We briefly recollect the technique employed by Singh and Roy \cite{sr70} for obtaining bound ${{d\sigma^{inel}}\over{d\Omega}}$. We collect
the necessary formulae to be useful in the sequel.\\
(1). The elastic differential cross section is expressed in terms of partial wave expansions 
\bea
\label{sec3.14}
 {{d\sigma^{el}}\over{d\Omega}}=\sum_{l=0}^{l=\infty}(2l+1)|f_l(s)^{el}|^2
 \eea
An upper bound is to be derived with following conditions which are obtained already from axiomatic field theory.  The conditions are\\
(I)
\bea
\label{sec3.15}
\sigma_t(s)={1\over s}\sum_{l=0}^{l=L}(2l+1){\rm Im}~f_l(s)^{el}P_l(cos\theta=0)
\eea
$L\sim {\sqrt s}log~s$ \\
(II)  $|F(s,t)|^2$ satisfies following inequality in the physical region of $cos\theta$
\bea
\label{sec3.16}
|F(s,t)|^2 &&\le \sum_{l=0}^{l=L}(2l+1)|f_l(s)^{el}|^2|P_l(1+{{2t}\over{{\bf P}_1^2}})|\nonumber\\&&
\le\sum_{l=0}^{l=L}(2l+1){\rm Im}~f_l(s)^{el}|P_l(1+{{2t}\over{{\bf P}_1^2}})|<s^N, N\le 2
\eea
It is known from convergence of partial wave expansion,
 inside Lehmann-Martin ellipse, because $cos\theta_0>1$, there is some cutoff value of $l=L$,
$L\sim {\sqrt s}log~s$, 
beyond which sum of all partial wave amplitudes, when we obtain expression for $\sigma_t(s)$, beyond $l>L_0$ is subdominant.
The problem of deriving the desired upper bound is reduced to using (I) and (II) as constraints. This is accomplished by appealing to Lagrangian
multiplier method. Thus, the upper bound on differential cross section is expressed in terms of $\sigma_t^{el}(s) $, \\
and (III)
\bea
\label{sec3.17}
{{\sqrt s}\over{R_M}}\sum_{l=0}^{\infty}(2l+1)|f_l(s)^{el}|^2P_l(1+{{t_0-\epsilon}\over{{\bf P}_1}^2})\le {\rm Im}~F(s, t_0-\epsilon)<s^2
\eea
Here $t_0$ is the right extremity of the LLE. 
Note that the argument of $P_l$  is greater than $1$ hence we did not write $|P_l(x)|$ in the above equation since $x>1$.
Thus $|f_l(s)|$ was bounded and the bound involved $P_L^2(cos\theta)$ where  $-1\le cos\theta\le +1$ and square of its derivative of Legendre
polynomial. 

For the inelastic case, (Singh and Roy studied some hadronic reactions also) we have derived inequalities for inelastic partial wave amplitudes from unitarity
considerations. Moreover, the inelastic total cross section, $\sigma_t(s)^{ab\rightarrow cd}$ gives an analogous expression as (\ref{sec3.14}).
However, there is no optical theorem for inelastic reaction.  Our results are, however,  not as strong as those of \cite{sr70}.
Note that $t_0=4m_{\pi}^2$ for most hadronic processes \cite{sr70}.
The  inequality that follows from unitarity condition is
\bea
\label{sec3.18}
|f_l(s)^{ab\rightarrow cd}|^2\le{\rm Im}~f_l(s)^{ab\rightarrow ab}
\eea
and inelastic total cross section is bounded as
\bea
\label{sec3.18a}
\sigma_t^{ab\rightarrow cd}\le {{4\pi}\over{T_0}}(ln{{s}\over{s_0}})^2
\eea
Here $\sigma_t^{ab\rightarrow cd}$ contains undetermined parameters
$T_0$ and $s_0$. We have explained why we cannot provide a value of $T_0$.
Moreover, the polynomial boundedness of imaginary part of inelastic process 
is indirectly used from polynomial boundedness of absorptive part
elastic amplitude (which is the Jin-Martin bound \cite{jin-martin}).  
We get a bound nevertheless for differential cross sections at 
high energies in forward
directions i.e. $\theta=0$ and in nonforward directions, $\theta\ne 0,\pi$ as given below.\\
(i) Forward scattering:
\bea
\label{sec3.19}
{{d\sigma^{ab\rightarrow cd}}\over{d\Omega}}|_{\theta=0}\le 
({{s}\over{16\pi T_0^2}})[log{{s}\over s_0}]^4
\eea
Note that $T_0$ corresponds to threshold mass of $t$-channel reaction 
($=4m_{\pi}^2$ usual hadronic reaction ):  
for $a+{\bar d}\rightarrow {\bar b}+c$
in {\it our hadronic} reactions.   Moreover,
differential cross section shows a peak in the forward direction.\\
(ii) Nonforward scattering, $\theta\ne 0 ~or~ \pi$
\bea
\label{sec3.20}
{{d\sigma^{ab\rightarrow cd}}\over{d\Omega}}\le 
({{1}\over{4\pi T_0^{3/2}}}){{1}\over{{\sqrt s}sin\theta}}[log{{s}\over s_0}]^3
\eea
For small angles, we may approximate $sin\theta\approx \theta$.
We discuss another type of inelastic reaction which might be of interests. Let us consider a reaction where only a single particle is observed
experimentally. Thus the process, known as {\it inclusive reaction} is $a+b\rightarrow c+X$ where $X$ is a single particle or multiparticle state and
is not subject to observation. Feynman \cite{feynman}  coined this word in the context of parton model to
analyze attributes of hadronic scattering. He argued that inclusive cross sections will exhibit scaling phenomena  in various kinematical regions . Those
feature were experimentally observed. The most familiar example of an inclusive reaction is deep inelastic $e+p$ scattering where energy of the scattered
electron is measured at different angles. The structure functions exhibit Feynman scaling. Our goal is not to discuss inclusive reaction and scaling
of its inclusive cross sections.  It is worthwhile to note that inclusive cross section measurements have 
some advantages over exclusive reactions such
as an inelastic process like $a+b\rightarrow c+d$ since inclusive reactions have several final state channels available. Moreover, if there is
conservation of a charge, when we search for such a charged particle, $c$, the unobserved inclusive state, $X$, must carry equal and 
opposite charge of the   experimentally observed particle, $c$.  Another point deserves mention here. For two body scattering, the total cross section is $a+b\rightarrow X$
is related to the imaginary part of the forward elastic cross section, 
${\rm Im}~F(s,t=0)^{ab\rightarrow ab}$, from the optical theorem.

 A generalized optical theorem was
derived by Mueller \cite{mueller}  for inclusive reaction. The differential cross section for $a+b\rightarrow c+X$ is related to 
the imaginary part of a forward scattering amplitude, ${\rm Im}~F^{(a+b+{\bar c} )\rightarrow (a+b+{\bar c)}}(s,M_X^2,t)$, where $\bar c$ is the antiparticle
of $c$. The forward six point amplitude depends on three kinematical variables $s$, $M_X^2$ and $t$ where $t$ is the momentum transfer squared
between particle $a$ and $c$ i.e. $t=(p_a-p_c)^2$. Note that the integration of the differential cross section 
${(E_c{{d\sigma^{ab\rightarrow cX}}\over{d^3p_c}}})$ is not just $\sigma_t(s)$ but it is $\sigma_t(s)<n_c>$ where $<n_c>$ is the average multiplicity
of outgoing particle $c$. We may consider the inclusive reaction to be a two body inelastic reaction, $a+b\rightarrow c+X$ where $X$ could be a single particle or
collection of many particles where $p_X$ is sum of  four momentum of all unobserved outgoing particles. Therefore, $p_a+p_b=p_c+p_X$ is the energy momentum
conserving relation and $t=(p_a-p_c)^2$. These kinematical aspects were 
considered by \cite{russian1,russian2}. Moreover, they extended their arguments
for inelastic two body collision case to the case of inclusive reaction. 
However, as noted in preceding discussions, these authors \cite{russian1, russian2} did
not take into account  the delicate issues associated with unequal mass 
inelastic scattering and analyticity in the complex $t$ variable. 
We have discussed how to identify the domain of holomorphy
in complex $t$-plane for two body unequal mass  inelastic scattering earlier. To remind the reader, the real value of $t$ is bounded as $-t_M\le t\le -t_m$ and we have
given expression for $t_m$ which is not same as that for equal mass case. 
Second point to recall is the complex-$t$ domain is expressed as union and intersections
of various complex domains as noted in (\ref{sec3.12}) - see the expression for ${\cal D}_R^{inelastic}$. Those arguments go through for the inclusive reactions
as well.  

The inclusive forward differential cross section is
\bea
\label{sec3.21}
({{{d\sigma^{ab\rightarrow cX}}\over{d\Omega_c}}})_{\theta_c=0}\le ({{s}\over{ 4T_0^2}})[log({{s\over{s_0}}})]^4
\eea
Note that the bound is an improvement of  \cite{russian1,russian2} result since the prefactor is fixed; whereas there was an arbitrary constant 
as prefactor
in their analysis. Moreover, the prefactor can only be determined after invoking Martin's theorem. Our bound is weaker than that 
of \cite{sr70} since we have not utilized the Lagrange multiplier technique as there are no experimental data for our case. 
If we sum over all states $X$ then the bound for  differential cross section is
\bea
\label{sec3.22}
\sum_X {{{d\sigma}^{ac\rightarrow cX}}\over{d\Omega_c}}(s,cos\theta)
\le {{1}\over{\pi {T_0}^{3/2}}}{{\sqrt s}\over{sin\theta_c}}[log({{s}\over{s_0}})]^3<n_c(s)(\theta_c)>
\eea
where $<n_c(s)>$ is average multiplicity of particle $c$. We mention in passing that derivation of upper bound on average multiplicity in a general setting
has been discussed in \cite{km}. We note that inclusive reactions to detect production of 'exotic' states may be a 
more optimistic scenario.  

\bigskip

\section{ Analyticity Properties of Inelastic Scattering Amplitude for $D=5$, 
Decompactified Theory}

This section contains analyticity properties  for $D=5$ theory where all 
spatial dimensions are noncompact. In case of a theory where it has only
one field, in $D=4$, according to UED  hypothesis the five dimensional theory
also has only one field. 
Therefore, for a single specie of field in $D=5$, the four point amplitude
is an elastic amplitude. We shall recall the essential results for this 
reaction in the first place. In case of multiple species of fields in a $D=4$ theory,
the higher dimensional theory (as per UED hypothesis) will be endowed with 
those many fields. In this case inelastic
four point inelastic  amplitudes will be permitted; moreover, in general,
these will
be unequal mass scatterings. 
We allude to this aspect in some detail in this section.
 We recall that we discussed analyticity properties of inelastic amplitude 
rigorously in the previous section. Next, based on those properties we derived
bounds on differential cross section and inclusive cross sections. 
We recapitulate essential results from \cite{jmjmp} for $D=5$ theory with 
single type of
field. Let the 4-point amplitude be denoted by $F^5(k_1k_2,p_1p_2)$ where 
$(p_1,p_2)$ are momenta of incoming particles 
and the corresponding 'in' state is $|p_1p_2~in>$.  Similarly, 
$(k_1,k_2)$ are momenta of outgoing particles and 'out' state is $|k_1k_2~out>$.
We follow a convention where masses of particles, in $D=5$ are also denoted 
by $m_a,m_b,m_c$  etc. These masses becomes zero modes of
compactified $R^{3,1}\otimes S^1$ theory. Indeed, 
 if we appeal to UED hypothesis, 
the masses appearing in $D=5$ (noncompact) theory will correspond
to zero modes of compactified theory. Whenever, we need to specify amplitudes, 
distributions and other entities of interests in $D=5$ theory we shall
explicitly use subscripts/superscripts, $'5'$.  The  
expression for $F^5( k_1k_2, p_1p_2)$ is derived from LSZ reduction technique. 
We choose to reduce the state with momenta $(k_1,k_2)$. We refer
the reader to our paper \cite{jmjmp} for detail derivation of the amplitude 
following LSZ technique. The sequence of arguments are that the two fields in 
'out' state are reduced.  The operation of KG operators
on the interacting fields leads to $R$ product of two source currents  
associated with these two fields besides derivatives  of $\delta$-functions. 
The presence
of such delta functions do not affect analyticity properties of scattering 
amplitudes as is well known. We do not intend to repeat these details here. 
 For 
$D=5$ case, the amplitude assumes the following  form.
\bea
\label{sec6.1}
F^5(k_1k_2,p_1p_2)= -\int d^5xd^5ye^{i(k_1.x+k_2.y)}<0|RJ^5(x)_cJ^5(y)_d|p_1p_2>
\eea

Here $k_1$ and $k_2$ are reduced. $J_5$ is the source current and we have 
used subscripts $c$ and $d$ for later conveniences as will
be evident soon. In order to study analyticity properties, it is essential to 
define three distributions. This is recognized from the fact that KG operators
act on  R-product of interacting fields $R\phi_c(x)\phi_d(y)$. The operation 
of KG operators
on the interacting fields leads to $R$ product of two source currents  
associated with these two fields:
$(\Box_x+m_c^2)(\Box_y +m_d^2)R\phi_c(x)\phi_d(y)=RJ_c^5(x)J_d^5(y)$ and
in addition  finite number of derivatives  of $\delta$-functions are present 
on the right hand side of this expression as discussed earlier.  Here 
$\Box+m^2$ stands for
five dimensional KG operator. In a local field theory
only finite number of derivatives of $\delta$-function appear. Their presence 
does not 
affection analyticity properties of the amplitude for following reasons. 
Let us consider derivative of the $\delta$-functions(which is define in terms
coordinate space variables. Then polynomials of momenta will appear; however, 
scattering amplitude is a function of Lorentz invariant variables (Mandelstam
variables: $s,t$ and $u$. Thus polynomials in $s,t$ and $u$ will appear. 
With this argument, we qualitatively argue that scattering amplitudes are
polynomially bounded. The presence of
of such delta functions do not affect analyticity properties of scattering 
amplitudes as is well known. In such cases,  we write subtracted dispersion 
relations.
We define following matrix elements  VEV of R-products products of 
operators which are 
distributions  as has been discussed earlier. However, 
we derive
  spectral representation to study support properties without going into 
all details (see \cite{jmjmp}).  We shall recall the previous results for 
elastic
  amplitude and then mention how analyticity properties inelastic four 
point amplitude are analyzed.
\bea
\label{sec6.2}
F_R^5(q)=\int d^5ze^{iq.z}\theta(z_0)<P_fR[J_l^5(z/2), J_m^5(-z/2)]|P_i>
\eea
where $|P_i>$ and $|P_f>$  are arbrtrary momentum states.
\bea
\label{sec6.3}
F_A^5(q)=\int d^5ze^{iq.z}\theta(-z_0)<P_fR[J_l^5(z/2), J_m^5(-z/2)]|P_i>
\eea
and
\bea
\label{sec6.4}
F_C^5(q)=\int d^5ze^{iq.z}<P_fR[J_l^5(z/2), J_m^5(-z/2)]|P_i>
\eea
Moreover, note that
\bea
\label{sec6.5}
F_C^5(q)= F_R^5(q)-F_A^5(q)
\eea
Note also that, as we have noted earlier,  $F_C^5(q)=0$ when both $F_R^5(q)$ 
and $F_A^5(q)$ vanish. We introduce a complete set of states
$\sum_n|p_n,\alpha_n><p_n,\alpha_n|={\bf 1}$ and 
$\Sigma_{n'}|p_{n'}\beta_{n'}><p_{n'}\beta_{n'}|={\bf 1}$. Here $\alpha_n$ 
and $\beta_{n'}$ stand
for all the quantum numbers permitted according to conservation laws of the 
theory. Note that unlike the case of compactified theory
where intermediate states also include KK towers, here we have no such 
considerations. For the $D=5$ theory, the intermediate states,
$|p_n,\alpha_n>$ and $|\alpha_{n'}\beta_{n'}>$  are physical states,
i.e. $p_n \ge 0,~p^0_n\ge 0$ and $p_{n'}^2\ge 0,~p^0_{n'}\ge 0$. 
The spectral representation for $F_C^5$, after opening up of the current 
commutator consists of two terms, $A_s^5(q)$ and $A_u^5$ which
are absorptive parts of $s$-channel and $u$-channel amplitudes respectively. 
These are 
\bea
\label{ch6.5a}
2A_s^5=\Sigma_{n'}<P_f|J_m^5|(0)|)p_{n'}={1\over 2}(P_f+P_i)+q,\beta_{n'}>
<\beta_{n'},p_{n'}={1\over 2} (P_f+P_i)+q|J_l^5(0)|P_i>
\eea
and
\bea
\label{ch6.5b}
2A_u^5=\Sigma_n<P_f|J_l^5(0)|p_n={1\over 2}(P_f+P_i)-q,\alpha_n><\alpha_n,p_n={1\over 2} (P_f+P_i)-q|J_m^5(0)|P_i>
\eea
The arguments of Section 3 can be appropriately rephrased so that the
conditions $F_R^5(q)=0$ and $F_A^5(q)=0$ hold simultaneously in unphysical
regions. Thus $F_C^5(q)=0$ in this domain. The JLD representation for $D>4$ has
been obtained in \cite{jmjmp}. Finally, the existence of SLE and SLE has been
proved my us.

Now we summarize our results for determination of Lehmann ellipses  in $D=5$ 
dimensional theory for scattering of unequal mass particles. This 
analysis is relevant for the case where two or more species of fields are 
present i.e. the case of a UED model which would have field contents of $D=4$
and they are promoted to higher dimensions. A detail discussion of derivation 
of Lehmann ellipses have been presented by us in Section III of our paper
\cite{jmjmp}. We consider inelastic process $a+b\rightarrow c+d$ with $m_a=m_b$ and $m_c=m_d$. Furthermore, ${\bf P}_1$ and ${\bf K}_1$ represent,
respectively, the c.m. momenta of initial (incoming particles) and outgoing 
(final) particles. 
 The small Lehmann Ellipse is parametrized as
  SLE
 \bea
 \label{sec2.20}
 E_{SLE}^{ab\rightarrow cd}:~~: | t + (P_1 - K_1)^2+|t + (P_1+K_1)^2| < 4|{\bf P}_1|{\bf K}_1| x_0(s)
 \eea
 where $m_a=m_b$ and $m_c=m_d$, for the case at hand. Furthermore,
 \bea
 \label{sec2.21}
 x_0=\bigg[1+{{({\bf M}_a^2-m_a^2)({\bf M}_c^2-m_c)^2}\over{K_1(s)^2s}}\bigg]^{1/2}
 \eea
 We are considering a $D=5$ theory. Here ${\bf M}_a$ corresponds to  two or 
more particle states states which which have same quantum
 number as particle 'a'. A similar definition holds  ${\bf M}_c$. 
 We have used ${\bf M}_a={\bf M}_b$ and ${\bf M_c}={\bf M}_d$ in arriving 
at the above formula. Note that $x_0>1$. \\
 (ii) LLE
 \bea
 \label{se6.21}
 E_{LLE}^{ab\rightarrow cd}: | t + (P_1 - K_1)^2+|t + (P_1+K_1)^2| <4 {\bf Z}_L(s)
 \eea
 where ${\cal Z}_L(s)$ is given by
 \bea
 \label{sec6.22}
 {\bf Z}_L(s)=&&\bigg(\bigg[{{({\bf M}^2_a-m_a^2)^2}\over{s}}\bigg]
 \times\bigg(\bigg[{{({\bf M}_c^2-m_c^2)^2}\over{s}}\bigg]\bigg)^{1/2}\nonumber\\&&
  + \bigg(\bigg[{{({\bf M}^2_a-m_a^2)^2}\over{s}}+P_1(s)^2\bigg]^{1/2}\times
\bigg[{{({\bf M}_c^2-m_c^2)^2}\over{s}}+K_1(s)\bigg]^{1/2}\bigg)
 \eea
 Here ${\bf Z}_L(s) $ is a function of kinematical variables and masses which 
are associated with $D=5$ theory.
 We repeat that in case of a $D=4$  theory, with several field contents, 
when we adopt UED hypothesis all those fields are defined in $D=5$. Note
 that upon $S^1$ compactification,  when the spacetime geometry is 
$R^{3,1}\otimes S^1$, the zero modes of the manifold are identified
 with the field
 contents of $D=4$ theory which was promoted to $D=5$ theory. Thus, 
now the theory with $S^1$ compactification will have KK towers. In summary,
 we have identified the spectrum of UED model and we are able to discuss
 inelastic reactions in this scenario. The purpose of this investigation 
(in $D=5$)
 is point out that the spectral representation is nonperturbative in
 character. Thus it accounts for all quantum corrections as is the 
characteristic  of
 LSZ approach. It is pertinent to cite the example of derivation of 
Froissart bound and host of rigorous bounds in hadronic scattering. 
Historically,
 the rigorous proof of Froissart does not rely on any perturbation theoretic 
approach. Of all the parameters; there is one undermined parameter, $s_0$, 
in Froissart
 bound.  All other parameters are physically measured quantities.
 This  bound has been
 tested against 
experiment ( for $D=4$ case). The same argument holds for derivation 
 of bound on slope slope of diffraction peak and some of the lower bounds.
 
 It remains to prove crossing property of four point function. 
We shall present, for $D>4$,    a general argument which is only valid for 
four point amplitude. This
 result is adequate to prove analyticity of four point scattering amplitude. 
We may assign following momentum configurations, 
without loss of generality,
  to the four particles scattering process. Note that the expression for
 Lehmann ellipses is defined in terms of Lorentz invariant kinematical 
  variables.  We choose a frame where the momenta are
given following kinematical configurations.
   \bea
   \label{sec6.23}
   p_a=(p_a^0, 0,0,0,0), ~p_b=(p_b^0,p_b^1,0,0.0),~p_c=(p_c^0,p_c^1,p_c^2,0,0)~p_d=(p_d^0,p_d^1,p_d^2,p_d^3,0)
   \eea
 Note that, with the above  assignment of {\t five momenta} for 
the external particles, 
$a,b,c,d$, we work in a four dimensional submanifold of the five dimensional
 momentum space. The four point scattering amplitude is 
Lorentz invariant and it depends on two Mandelstam variables. Therefore, 
we are
 in a position to invoke theorem of Bros-Epstein-Glaser \cite{beg}  to prove 
 crossing for four point function.
We cannot prove crossing for $N$-point functions, $N>4$ in $D>$, by invoking
this argument; moreover, it remains a challenge to prove crossing for
the general case of N-point functions. 
 
 The analyticity properties of four point amplitude, for elastic processes, 
has been established rigorously for higher dimensional spacetime \cite{jmjmp}
 already. Moreover, analog of Martin's theorem was proved in the above article.
 Consequently fixed-$t$ dispersion relations have been written down for the 
amplitude when $t$ lies inside Lehmann-Martin ellipse. 
We argue that our results on analyticity properties of scattering amplitude 
derived in Sections 3 and 4 will go through \cite{jmjmp}. 
However, there is a distinction. 
If
 we have a single field in $D=5$ then the four point amplitude corresponds 
to elastic process. Thus inessential complication for inelastic reaction do not
 appear.
 On the other hand if $D=5$ theory is endowed with several fields in accordance
 with UED hypothesis then four point amplitude describes elastic
 as well inelastic reaction; indeed allows provisions for unequal mass 
scattering. Note that for 4-point amplitude, describing inelastic reactions 
in $D=5$,
 is not so complicated when we study its analyticity. The reason is that, 
as we have argued, it is possible to reduce the problem to scattering 
in $D=4$
 spacetime which is special to four point amplitude. In view of preceding
 remarks, we can use the results derived in previous sections. However, there
 are certain differences which we shall point out at the appropriate juncture.
 
 We can write down  fixed-$t$  unsubtracted dispersion relations; if the 
dispersion integrals do not converge we can implement subtraction which
will have finite number of terms.
 We recall that
 the amplitude is polynomially bounded in $s$ (see \cite{jmjmp}).
 Thus the dispersion relation is  
\bea
\label{sec6.24}
F_5(s,t)^{(ab\rightarrow cd)}={1\over{\pi}}\int_{s_{thr}}^{\infty}{{ds'{{\rm Im}~F_5^{(ab\rightarrow cd)}(s,t)}}\over{(s'-s)}}+
{1\over{\pi}}\int_{u_{thr}}^{\infty}{{du'
{{\rm Im }~F_5^{({{a{\bar c}}\rightarrow d{\bar b}})}(u',t)}}\over{(u'-u})}
\eea

Now we proceed to discuss bounds on near forward amplitude and near forward 
differential cross section at asymptotic energies. 
We consider a five dimensional
theory which accommodates several species of field; thus inelastic 
reactions are allowed. The case of a single scalar field has been investigated
thoroughly in \cite{jmjmp}. The,
domain of $t$-plain analyticity is the Large Lehmenn Ellipse (LLE).
    Let us recall that the amplitude is expanded
in partial waves with Geggenbauer polynomials,
$C_l^{\lambda}(1+{{2t}\over s})$,  as basis functions \cite{jmjmp}.
\bea
\label{sec6.25}
F_5^{(\lambda, ab\rightarrow cd)}(s,t)={\rm Const}~\Sigma_{l=0}^{\infty}(l+\lambda)f_l^{(\lambda, ab\rightarrow cd)}(s)C^{\lambda}_l(1+{{2t}\over s})
\eea
where $\lambda={1\over 2}(D-3)$, D is number of spacetime dimensions and 
$\lambda=1$ for $D=5$.  
Note that, for inelastic reactions, we are unable to
utilize positivity properties of partial wave amplitudes,  
$f_l^{\lambda ab\rightarrow cd}$. We have derived a bound 
\cite{jmjmp15} on $F^{\lambda}(s,t)$ for $|t|<T_0$ where
$T_0$ is right extremity of  LLE.
\bea
\label{ch6.26}
|F^{(\lambda, ab\rightarrow cd)}(s,t)|\le A_2({{1}\over{|t|}})^{1/2(1+\lambda)}
({{1}\over{T_0}})^{\lambda /2}
s^{(1+(N-1))
{\sqrt{{|t|}\over{T_0}}}}({\rm }{logs/ {s_0}})^{\lambda}
\eea 
Here $A_2$ is a constant, independent of $s,t$; $s_0$, a dimensionful
constant is introduced to make argument of $log$ dimensionless
in the above equation. $s_0$ cannot be determined from first principle (as
is the case in $D=4$ for Froissart bound). The appearance of $T_0$
and its detrmination deserves some discussion. It is the right extremity of
the LLE. Intuitively, $T_0$ is related to the lowest two particle threshold
in the $t$-channel i.e. mass squared of lightest two particle state. For
$D=4$, in the case of hadronic collision, $T_0=t_0= 4m_{\pi}^2$, first 
determined by Martin. In case of production of exotic states in 
$D=5$ it will be threshold for $t$-channel cut. Thus value of $T_0$ 
cannot be determined. So is the case for $S^1$ compactified theory
in four dimensions. $T_0$  is an independent free parameter and 
it determined from the value of threshold for the crossed $t$-channel 
reaction: $p+{\bar K}_1\rightarrow p+{\bar K}_2$ from the direct channel 
process $p+K_1\rightarrow p+K_2$. Here ${\bar K}_1, {\bar K}_2$ are 
respectively anti-particles of $K_1$ and $K_2$. Let take $m_p\approx 1$ GeV
and $m_{K_1}\approx 1$ TeV equal to mass of $m_{K_2}$ as a ball park range.
We note that the $t$-channel cut begings  at $t_{th}= 4m_{K^2}$, if we ignore
mass of proton compared to $m_K$ which is order of TeV in LRC scheme. We have
discussed the important role of $t_{th}$ in derivation of Lehmann ellepses.
Therefore,$T_0=4m_K^2$ which is the analog of $t_0=4m_{\pi}^2$ is strong
interactions, appearing as the prefactor of Froissart bound.  For the case
at hand $T_0$ is not determined as a number from experimental data.
Then $T_0=(m_p+m_{K_1})^2$ and is order of $1$ TeV. 
Moreover, $m_{K_2}^2\approx{{n}\over R}^2, n\in {\bf Z}$, 
where $R$ is compactification.
Therefore, the mass scale is related to compactification radius as is expected.
We have  introduced $s_0$ so that argument of $log$ is dimensionless; $s_0$ cannot be
determined from first principle.
We have emphasized  that due to polynomial boundedness and
convergence of partial wave expansion in a larger complex domain
the partial wave amplitudes (rather absorptive part) are
cut off beyond certain $L_0$ exponentially by  cut off  
by $log s$. We have proved $L-0$  is independent of $D$.  
  In order to derive bound on near forward inelastic
 scattering amplitude for the five 
dimensional theory, Note that forward scattering amplitude is bounded as
\bea
\label{ch6.27}
|F_5(s,t=0)|^2\le {\rm const}~({{1}\over{2{\sqrt R_M}}})^3
(log~{{s}\over{s_0}})^3(1+
({R_M/Const})^2)^{-1/2}
\eea
The inelastic total cross section is bounded as 
\bea
\label{ch6.27a}
\sigma_{5t}^{ab\rightarrow cd}\le {{4\pi}\over{T_0}}(log{{s}\over{s_0}})^3
\eea
Here $R_M$ is the radius of the Martin's circle.
Note that this bound is weaker one in the sense that it is $(log~s)^3$ 
compared to compactified theory in $D=4$ where cross section is bounded by  
$(log{{s}\over{s_0}})^2$.  Thus,
in what follows, from a phenomenological perspective in the accessible 
LHC energies the bound for $D=5$ theory is higher by a factor of 
$log{{s}\over{s_0}}$. This
'weaker' bound trend continues to appear for other measurable parameters. 
If the data exhibits violations of $D=4$ bounds (for compactified case) 
there will
be reasons to argue that an extra dimension is decompactified.

For $D=5$, diffrential cross section is bounded as ($\theta$ small)
\bea
\label{ch6.29}
{{d\sigma}\over{d\Omega}}\le { {{\rm const}}\over{T_0^{3/2}{\sqrt s}\theta}}
(log{{s}\over{s_0}})^4
\eea
We note that the differential cross section, ${{d\sigma}\over{d\Omega}}$, 
after   approxmating
Gegenbauer polynomial for small scattering angle in near forward direction,
  has a weaker bound in asymptotic energies.  The method
adopted in Section 4, for $D=4$ ( with $S^1$ compactification) is applicable to derive similar to bounds in case  $D=5$ theory. Therefore,
we shall not repeat those computations here. These bounds will also 
be weaker compared to those of Section 4 so far as the energy dependence
is concerned.

\bigskip

\begin{figure}[!htbp]
  \begin{center}
  \includegraphics[width=0.7\textwidth]{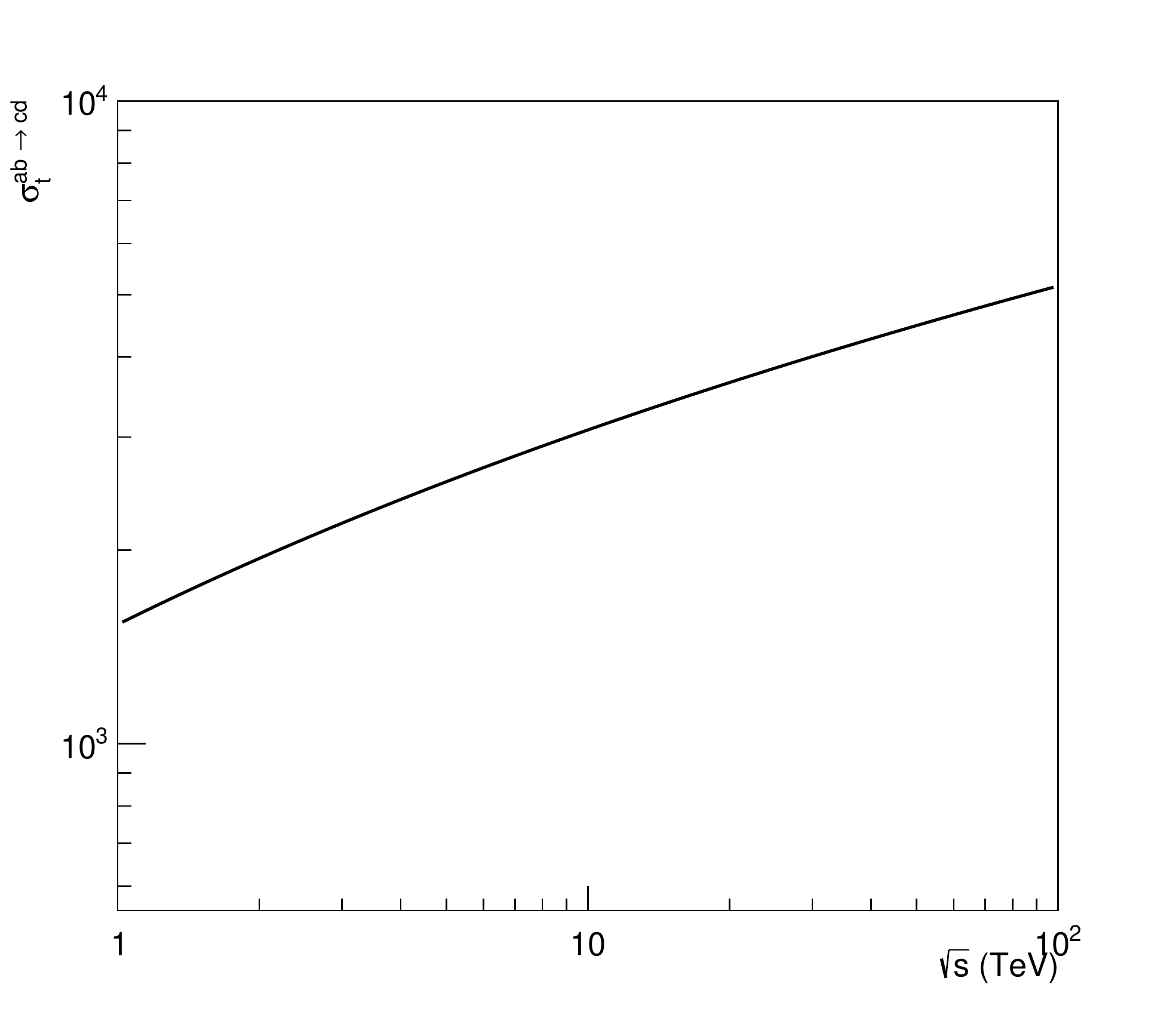}
  \caption{ Depicts bound $\sigma_t^{ab\rightarrow cd}$ for the $S^1$ compactified theory. It is plotted against energy.}
  \label{fig:bound1}
  \end{center}
\end{figure}

\begin{figure}[!htbp]
  \begin{center}
  \includegraphics[width=0.7\textwidth]{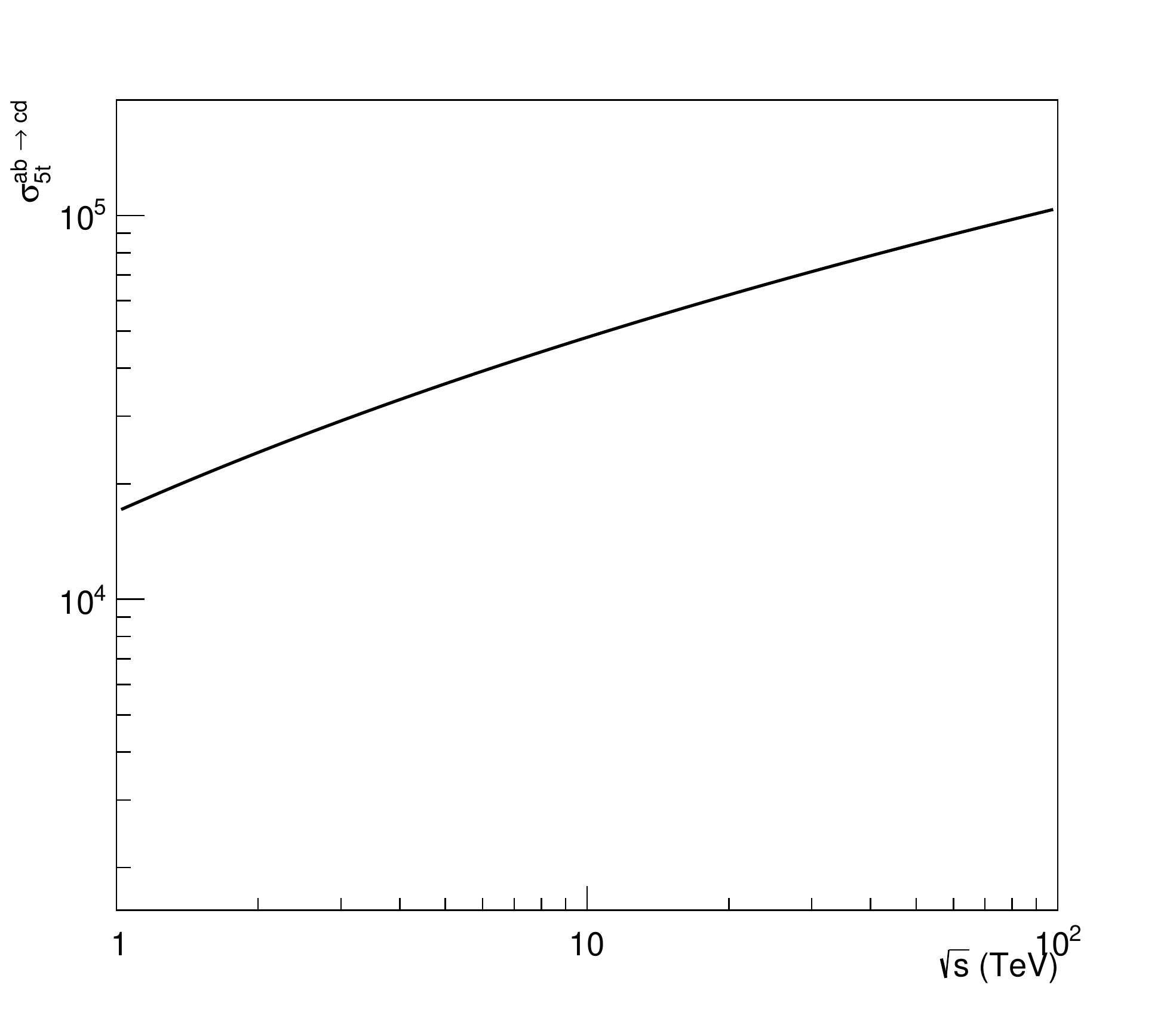}
  \caption{ This shows the bound for inelastic cross section, $\sigma_{t5}^{ab\rightarrow cd}$ for decompactified theory.}
  \label{fig:bound1}
  \end{center}
\end{figure}

\begin{figure}[!htbp]
  \begin{center}
  \includegraphics[width=0.7\textwidth]{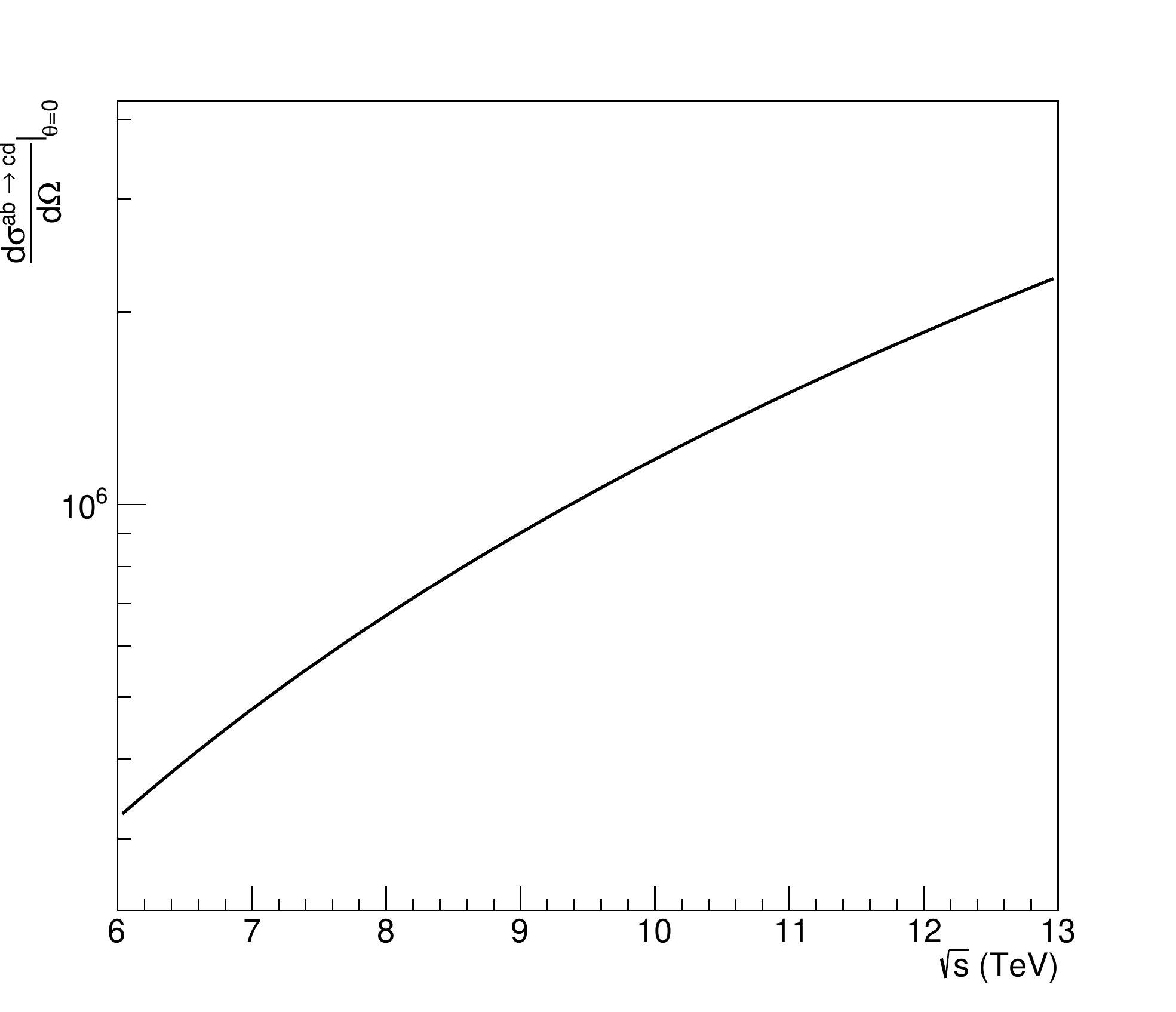}
  \caption{ This shows bound for forward differential cross section for the $S^1$ compactified theory.}
  \label{fig:bound1}
  \end{center}
\end{figure}

\begin{figure}[!htbp]
  \begin{center}
  \includegraphics[width=0.7\textwidth]{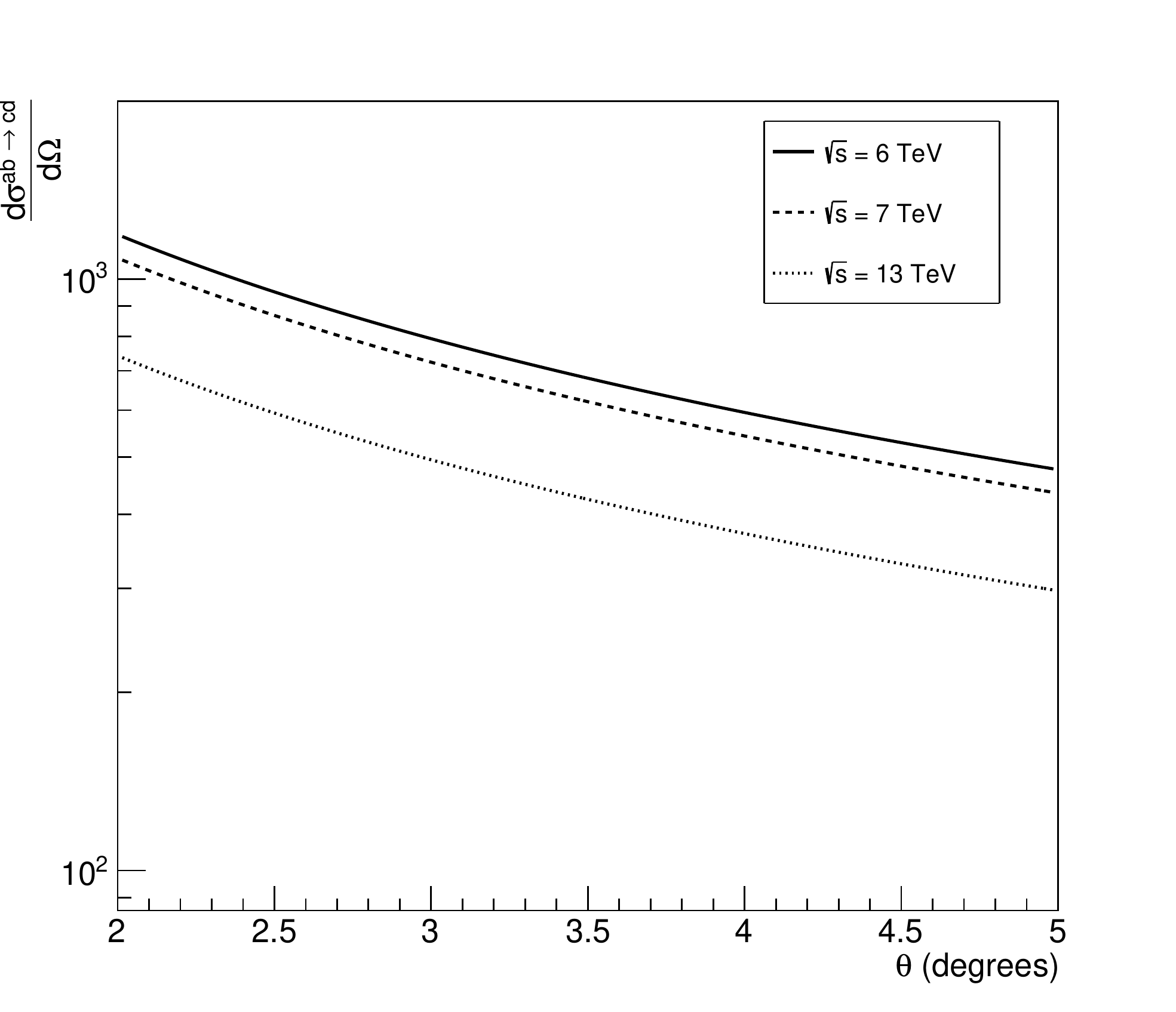}
  \caption{Depicts bounds on near forward differential cross sections for $ab\rightarrow cd$ at angles $\theta=2^0,3^0$ and $\theta=5^0$.}
  \label{fig:bound1}
  \end{center}
\end{figure}

\section{ Summary and Discussions}

\bigskip
We considered production of Kalulza-Klein states, arising from $S^1$ 
compactification of a $D=5$ theory. In the LRC scenario,
the excited KK states  might have masses in the range of TeV 
and might be discovered at CERN-LHC 
collider. We have discussed some aspects of the production
of the exhotic states in pp collision in the introduction section.
We have assumed that KK charge is conserved in the collision of high
energy protons and consequently,
 final two body KK states carry equal and opposite
KK charge. Moreover, baryon number is conserved. There are several phenomenological  models which envisage
detail properties of  the idea of LRC paradigm  and suggest that 
experimentalists might succeed in observing such states.  
We derived bounds on cross section starting from general principles of 
relativistic
quantum field theories.  These bounds might be useful to them.
We envisaged a scenario where the theory 
does not have any compact spatial dimension. In  other words,
all spatial dimensions are noncompact and the geometry is $R^{4,1}$. 
One of the advantages of our approach is that the results do 
not depend on detail inputs of a model. We had to
overcome several obstacles in deriving the results. First of all, 
as we have presented in detail, the case of unequal mass inelastic
two body scattering process. It was argued that such scatterings have  to be treated carefully.
 We obtained the desired analyticity properties of inelastic amplitude 
in Section 4.
It is worthwhile to note that some of the issues, which were overlooked in previous investigations, have been addressed in this article and have
been resolved.  We employed refined techniques to this end. We have noted that
in the previous works \cite{russian1,russian2} arbitrary constants appeared in the expressions for bounds and some weak
assumptions were incorporated. On this occasion, since
issues related to analyticity domains of amplitude have been resolved, those arbitrary constants have been determined. Now more refined
bounds on forward differential and nonforward differential cross sections have been presented here. We have considered 'inclusive'
processes for our purpose and we have obtained various bounds. We utilized an old theorem, derived for average multiplicity, to argue
that the resulting bound for integrated inclusive reaction could  be improved.
  It is important to mention that the energy scale, $s_0$ which 
invariably appears
in the derivation of these bounds  in the power of 
logarithms in the form of $log{{s}\over{s_0}}$, is not fixed. 
Even, in the derivation of Froissart-Martin bound, for hadronic colllsions, 
the dimensionful constant, $s_0$, 
remains undetermined starting from axioms of field
theory whereas $t_0$ is determined. Prudently, $s_0$ may be guessed 
  to 
be order of $1$ $GeV^2$ relating to mass of proton.
 Therefore, the question arises what value of $s_0$ is to be adopted
for our situation? 
A plausible answer is
that it should be the scale of compactification.  However, so far there is 
no experimentally confirmed evidence for large-radius-compactification
hypothesis. The positive aspect is that the energy dependence of bounds 
have been determined from first principles in the frameworks of local quantum
field theories.

Petrov \cite{petrov} has advanced argument that a $D>4$ theory below 
decompactification scale will exhibit attributes of a $D=4$ theory
 with specific model calculations. We have already shown that a 5-dimensional 
theory
with 
$S^1$ compactification is endowed with analyticity
properties of a $D=4$ theory in a rigorous framework. We confirm
conjecture of Petrov that the compactified theory leads to a bound
on total cross sections same as the Froissart bound proved for
a $D=4$ theory. We have 
 proved fixed-$t$ dispersion relations for elastic processes. 
This article deals
with study of analyticity properties of scattering amplitude for 
inelastic reactions. 
Therefore, the production of 'exotic'   state were dealt with
in our new results. Moreover,
we have derived bounds on production cross sections in 
forward direction and in small 
finite scattering angles. Therefore, our results will be useful
for experiments currently going on at LHC. We have also studied asymptotic 
behaviors of amplitudes for a decompactified, $D=5$, theory.
We have proposed to test presence of extra spatial dimensions in 
high energy collisions in the accessible energy range. Let us imagine that
one extra spatial dimension is decompactified. We would like to 
ask is there a way to detect decompactification in high energy collisions? 
Our proposal
\cite{nm} is to analyse energy dependence of $\sigma_t$. The particle data group has presented the fit to hadron total cross sections, $\sigma_t$,
over a wide range of energies for several processes where the energy dependendence is assumed to be Froissart-Martin bound saturating i.e.
$\sigma_t\approx (log~{s\over{s_0}})^2$ with the energy scale fitted to be 
$s_0=16~GeV^2$. In case an extra spatial dimension is decompactified, there
is a possibility that data might exhibit deviation from Froissart-Martin 
bound. Nayak and Maharana \cite{nm} considered total 
cross section, $\sigma_t$ data for combined $pp$ and $p{\bar p}$
scatterings  and carried out a fit from ISR energy to Tevatron energy, 
and up to LHC energy. Furthermore,  data from Cosmic rays
were included in the analysis. 
We found no evidence for violation of $D=4$ Froissart bound. An optimistic
argument might be that an extra domension is decompactified at LHC energy
which does not saturate $D=5$ bound. However, our fit considered data from ISR
energy upwards and therefore, we do not believe  measurements of total
cross sections are  going to provide grounds for decompactication at LHC.

It has been suggested by Andr\'e Martin \cite{martin2018} that  if a  
new threshold opens up in a high energy scattering it will have 
important implications
for analyticity properties of amplitude. One of the implications is that  
the  amplitude might have a large real part.
In the context of present investigation,  when a pair of KK state is produced, 
it opens up a new threshold.  Consider a scenario when a pair of KK states
are produced, consistent with all conservation laws. At the threshold 
of production of a pair of exotic particles, the cross section for collision 
of
ordinary particles gets affected.  
 According to Martin, dispersion relations survive, furthermore,  
it is necessary to  put an extra term in total cross
section. This extra term has to be put at a high energy which  might upset 
 the 
real part at  lower energies below the new threshold when we write dispersion
relation. In view of above remarks,
if there is a threshold for production of KK states in LHC energy regime, 
measurements  of real part of scattering bellow threshold energies will provide 
indirect evidence for existence of new threshold. Of course, the real part measurements would not provide a clinching  evidence for production
of KK states; however, it might provide a possible lead to look for KK states.  


We have plotted our bounds as a function of energies and gone beyond
the existing LHC energies. We recall that $T_0$ or its fractional
power appear as prefactor in all the equations for the bounds. As we have
mentioned repeatedly, we cannot extract value of $T_0$ from experimental 
data. Therefore, while plotting the bounds, we have factored out $T_0$ (or its
powers from the expressions for bounds). In view of above mentioned 
remarks, the figures exhibit the energy bounds. 
Figure 1 shows the bound on $\sigma _t^{ab\rightarrow cd}$ 
for $S^1$ compactifies case as a function of energy.  
We have plotted in figure 2,
 bound for $\sigma_{t5}^{ab\rightarrow cd}$, for the five dimensional
theory. Figure 3 shows the bound for inelastic 
differential cross section in forward
direction
 for the compactified $d=5$ theory. Figure 4 depicts bounds on differential
cross section in near forward direction for the compactified theory,
$R^{3,1}$ at angles  $\theta=2^0,3^0$ and $5^0$ and note 
that
differential cross section falls of rapidly as $\theta$ increases.
This
bound might interest experimentalists who are searching for exhotic
states at LHC.
 
We have refrained from presenting a figure for bound on differential cross 
sections for the decompactified theory since this bound is weaker compared to 
the differential cross section for compactified theory.   
 
We end with following closing remarks. Note that  
 the results presented here are not specific to LRC proposal alone. 
LHC is also searching for other exotic states besides KK states,
such as SUSY particles. The bounds presented
here do not specifically utilize ingredients of LRC models. Therefore, 
they hold good for production of other particles in inelastic reactions. 
For example,
there is a high expectation that supersymmetric particle will be produced 
in LHC. The bounds derived here will play an useful role. 
If exotic particles are
discovered at CERN collider, their properties can be used to study several 
attributes  in a model independent manner. We remind
that, in the context of collisions of high energy hadron, 
the axiomatic results have played important roles to constrain models not 
only in high energy collisions but
have imposed restrictions in low energy physics like $\pi\pi$ and $\pi K$ 
reactions. Moreover, those bounds have been  experimentally tested
and violations have not been reported.  

\bigskip 
{\bf Acknowledgements}: I would like to thank Ignatios Antoniadis for
very useful discussions and valuable suggestion. I thank Aruna Kumar Nayak
and Sanjay Kumar Swain for their critical remarks and providing me some
information about search for states arising in LRC models at CERN.
I also thank Aruna Kumar Nayak  for his comments to improve the
manuscript. 
I thank  Matthias Gaberdiel for a very warm and gracious hospitality at 
the Institute of
Theoretical Physics of ETH. It is a pleasure to acknowledge Pauli Center 
at ETH, Zurich, for  very cordial 
atmosphere  and for generous financial supports.

\bigskip

\end{document}